\documentclass[preprint]{article}
\usepackage{graphicx}
\usepackage[latin1]{inputenc}
\newcommand{\sca}[2]{\left\langle #1 | #2 \right \rangle}

\newcommand{\bfB}{{\bf B}}
\usepackage{amsmath}
\usepackage{amssymb}
\usepackage{rotating}
\begin{document}

\title{On the magnetic fields generated by experimental dynamos}

\author{F. P\'etr\'elis \thanks{Corresponding
author. Email: petrelis@lps.ens.fr}, N. Mordant and S. Fauve
\\\vspace{6pt} Laboratoire de Physique Statistique , CNRS UMR 8550,\\ Ecole Normale Sup\'erieure, 24 rue  
Lhomond, 75005 Paris France}
\date{}
\maketitle

\begin{abstract}
We review the results obtained by three successful fluid dynamo  
experiments and discuss what has been learnt from them about the  
effect of turbulence on the dynamo threshold and saturation. We then  
discuss several questions that are still open and propose experiments  
that could be performed to answer some of them.
\end{abstract}

\section{Fluid dynamos}

It is now believed that magnetic fields of planets and stars are  
generated by the motion of electrically conducting fluids through the  
dynamo process. This has been first proposed by Larmor (Larmor 1919)  
for the magnetic field of the sun. Assuming the existence of an  
initial perturbation of magnetic field, he observed that ``internal  
motion induces an electric field acting on the moving matter: and if any
conducting path around the solar axis happens to be open, an electric  
current will flow round it, which may in turn increase the inducing  
magnetic field. In this way it is possible for the internal cyclic  
motion
to act after the manner of the cycle of a self-exciting dynamo, and  
maintain a permanent magnetic field from insignificant beginnings, at  
the expense of some of the energy of the internal circulation" (for  
reviews of the subject, see Moffatt 1978, Zeldovich, Ruzmaikin and  
Sokoloff 1983, Roberts 1994).

The minimum set of parameters involved in a fluid dynamo consists of
the size of the flow domain, $L 
$, the typical fluid velocity, $V$, the density,  $\rho$, the  
kinematic viscosity, $\nu$, the magnetic permeability of vacuum, $ 
\mu_0$, and the fluid electrical conductivity, $\sigma$. For most  
astrophysical objects, the global rotation rate, $\Omega$, also plays  
an important role. Three independent dimensionless parameters thus  
govern the problem.
We can choose the magnetic Reynolds number, $R_m$, the magnetic  
Prandtl number, $P_m$, and the Rossby number $Ro$

\begin{equation} R_m = \mu_0 \sigma V L, \;\;\; P_m = \mu_0 \sigma  
\nu, \;\;\; Ro = \frac{V}{L \Omega}.
\end{equation}
For planets and stars as well as for all liquid metals in the  
laboratory, the magnetic Prandtl number is very small, $P_m < 10^{-5} 
$. Magnetic field self-generation can be obtained only for large  
enough values of $R_m$ for which Joule dissipation can be overcome  
(for most known fluid dynamos, the dynamo threshold $R_{mc}$ is
roughly in the range $10-100$). Therefore, the kinetic Reynolds  
number, $Re = VL/\nu = R_m/P_m$, is
very large and the flow is strongly turbulent.  This is of course the  
case of planets and stars which involve huge values of $Re$ but is  
also true for dynamo experiments with liquid metals for which $Re >  
10^{5}$.
Direct numerical simulations are only possible for values of $P_m$  
orders of magnitude larger that the realistic ones for the sun, the  
Earth or laboratory experiments.  First because it is not possible to  
handle a too large difference between the time scale of diffusion of  
the magnetic field and the one of momentum; second, a small $P_m$  
dynamo occurs for large $Re$ and requires the resolution of the  
small spatial scales generated by turbulence.  Strongly developed  
turbulence has also some cost for the experimentalist. Indeed, the  
power needed to drive a turbulent flow scales like $P \propto \rho  
L^2 V^3$ and
we have
\begin{equation} R_m \propto \mu_0 \sigma
\left(\frac{PL}{
\rho}\right)^{1/3}.\end{equation} This formula has simple  
consequences: first, taking liquid sodium (the liquid metal with the  
highest electric conductivity),
$\mu_0 \sigma \approx 10 \,{\rm m}^{-2} {\rm s}$, $\rho
\approx 10^3 \,{\rm kg} \, {\rm m}^{-3}$,
and with a typical length scale
$L \approx
1\, {\rm m}$, we get $P \approx R_m^3$; thus a mechanical power larger
than $100\,
{\rm kW}$ is needed to reach a dynamo threshold of the order of $50$.
Second, it appears unlikely to ever operate experimental dynamos at  
$R_m$ large compared with $R_{mc}$. Indeed, it costs $8$ times more  
power to reach $2 R_{mc}$ than to reach the dynamo threshold. In  
conclusion, most
experimental dynamos should have the following characteristics:
\begin{itemize}
\item they bifurcate from a strongly turbulent flow regime,
\item they operate in the vicinity of their bifurcation threshold.  
\end{itemize}

Although the values of $R_m$ 	and $P_m$ that can be obtained in  
laboratory experiments using liquid sodium are not too far from the  
ones of the Earth core, it would be very difficult to perform  
experiments with large $R_m$ at $Ro$ significantly smaller than unity  
whereas we have $Ro \approx 10^{-6}$ for the Earth core. The  
comparison is of course also difficult in the case of the sun:  
although $Ro$ is of order one for the solar convection zone, $R_m$ is  
more than six orders of magnitude larger than in any laboratory  
experiment. As said above, the situation is worse when direct  
numerical simulations are considered. We thus cannot claim that  
cosmic magnetic fields can be reproduced at the laboratory scale  
except if we can show that the dynamics of the magnetic field weakly  
depends on some dimensionless parameters.
\begin{table}
Relevant dimensionless parameters for some planets and the  
sun (from Zeldovich {\it et al.} 1983)
   and for laboratory fluid dynamos. ($R_m$ has been evaluated on the  
full size even in the case of scale separation).
{\begin{tabular}{@{}lcccccc}
    $$
   & $P_m$ & $R_m$
          & $Re$
          & $Ro^{-1}$ &   \\
    Earth &$5 \times 10^{-7}$& $500$ & $10^9$ & $10^{6}$
\\
    Jupiter &$10^{-6}$& $10^{6}$ & $10^{12}$ & $10^{6}$
    \\
   Sun & $5 \times 10^{-8}$ & $10^8$ & $2 \times 10^{15}$ & $1$
          \\
   Experiments & $10^{-5}$ & $50$ & $10^5 - 10^7$ & $0 - 1$
   \\

\end{tabular}}

\end{table}



As already mentioned, laboratory dynamos operate in the vicinity of the instability threshold but at very high values of the Reynolds number. This give rise to a very interesting example of instability that differs in many respects from usual hydrodynamical instabilities. The dynamo bifurcation occurs from a base state which is fully turbulent.  This may play a role on various aspects of the dynamo process and it raises several questions. 

\begin{itemize}

\item
What is the effect of turbulent velocity fluctuations on the dynamo onset? Will they favor or inhibit dynamo action? Can they change the nature of the bifurcation?

\item
Above onset, at which amplitude does the magnetic field saturate?
Does it display an anomalous scaling with respect to the distance to threshold due to turbulent fluctuations?

\item
What are the  statistical properties of the fluctuations of the magnetic field?
  
\end{itemize}  

We will discuss this problem in connection with existing laboratory dynamo experiments. The first ones, performed in Karlsruhe  and Riga, have been designed by taking into account the mean flow alone. Turbulent fluctuations have been inhibited as much as possible by a proper choice of boundary conditions. On the contrary, the VKS experiment has been first motivated by the study of the possible effects of turbulence on the dynamo instability. 

The paper is organised as follows: in section 2, we recall the results of Karlsruhe and Riga experiments. Section 3 concerns the effect of small turbulent fluctuations on the dynamo threshold and emphasises
the effect of the Reynolds number on the scaling law for the  
saturated magnetic field above
threshold. Section 4 considers some questions related to dynamos  
generated by strongly
turbulent flows, i. e. with less geometrical constraints than the
Karlsruhe and Riga experiments. The results of the VKS experiment  
with counter-rotating impellers are presented in section 5. A  
possible dynamo mechanism for the VKS experiment together with some  
additional comments are presented in section 6. Finally, some open  
questions and other possible dynamo experiments are discussed in  
section 7.

\section{The Karlsruhe and Riga experiments}

The first homogeneous fluid dynamos have been operated in liquid  
sodium in Karlsruhe
(Stieglitz and M\"uller 2001) using a flow in an array of pipes set-up in
order to mimic a spatially periodic flow proposed by G. O. Roberts  
(1972),
and in Riga (Gailitis {\it et al.} 2001) using a Ponomarenko-type flow
(Ponomarenko  1973). We first recall the flow geometries and briefly  
review the results obtained by both groups.

\begin{figure}[!htb]
\vspace{0.0 in}
\begin{center}
\includegraphics[width=5 in]{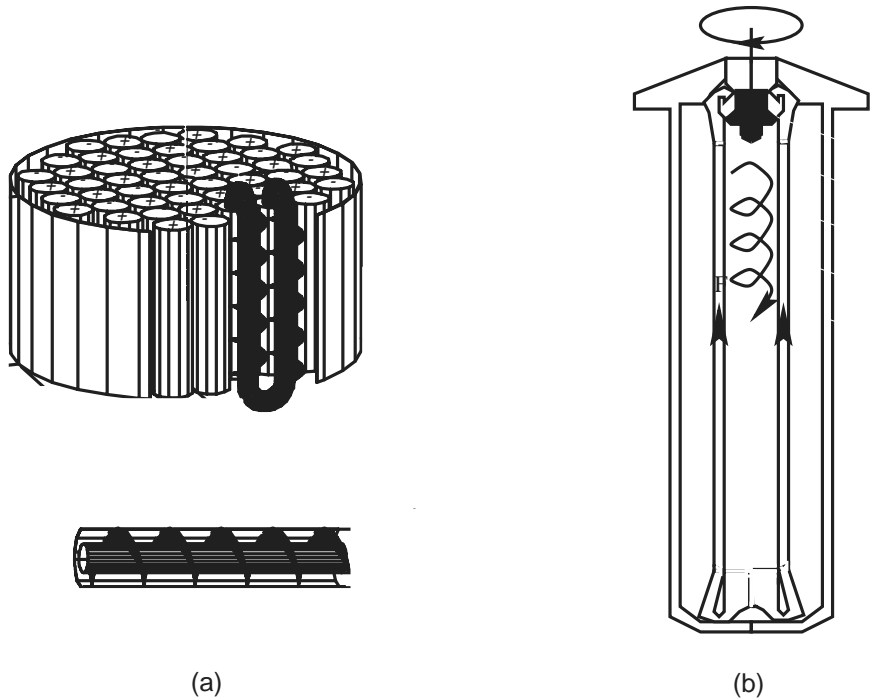}
\end{center}
\caption{Schematics for the experiments from Karlsruhe (a) and Riga  
(b) which
show how helical flow is forced by guiding the sodium through steel
channels (from Stieglitz and M\"uller 2001 and Gailitis {\it et al.} 2001).}
\label{karlsruheriga}
\end{figure}

\subsection{The Karlsruhe experiment}

The experiment in Karlsruhe, Germany, was motivated by a kinematic
dynamo model developed by G.O. Roberts (Roberts 1972) who showed  
that various periodic flows can generate a magnetic field at large
scale compared to the flow spatial periodicity. One of the cellular flows he 
considered is a periodic array of vortices with the same  
helicity. Flows with such topology drive an $\alpha$-effect that can  
lead to dynamo action. This mechanism is quite  
efficient at self-generation (in the sense of generating a magnetic field 
at a low magnetic Reynolds number based on the wavelength of the flow). 

A dynamo based on this mechanism was constructed and run with 
success  in Karlsruhe.
A sketch of the experiment is shown in figure \ref{karlsruheriga}a.   
The flow is located in a cylindrical vessel of width $1.85\, {\rm  m}$ and  
height $H=0.7\, {\rm  m}$. It contains $52$ elementary cells placed on a  
square lattice.  Each cell is made of two coaxial pipes: an helical  
baffle drives the helical flow in the outer cylindrical shell whereas  
the flow in the inner shell is axial. In two neighbouring cells, the  
velocities are opposite such that the helicity has the same sign in  
all the cells. Although the volume is finite instead of the infinite  
extension assumed by G. O. Roberts, the dynamo capability of the flow  
is not strongly affected  in the limit of scale separation, i.e.,  
when the size $L$ of the full volume is large compared to the  
wavelength $l$ of the flow (Busse {\it et al.} 1996). In this limit, the  
relevant magnetic Reynolds number involves the geometrical mean of  
the two scales as a length scale, and the geometrical mean of axial and  
azimuthal velocities as a velocity scale. However, it can be shown  
using simple arguments that it is not efficient to increase too much  
the scale separation if one wants to minimise the power needed to reach
the dynamo threshold (Fauve and P\'etr\'elis 2003).
The flow is driven by three electromagnetic pumps and the axial and  
azimuthal velocities are independently controlled. The liquid sodium  
temperature is maintained fixed by three steam-evaporation heat  
exchangers. Measurements of the magnetic field were made both locally  
with Hall-probes and globally using wire coils. Pressure drops in the  
pipe and local velocity measurements were also performed.

\begin{figure}[!htb]
\vspace{0.0 in}
\begin{center}
\includegraphics[width=4 in]{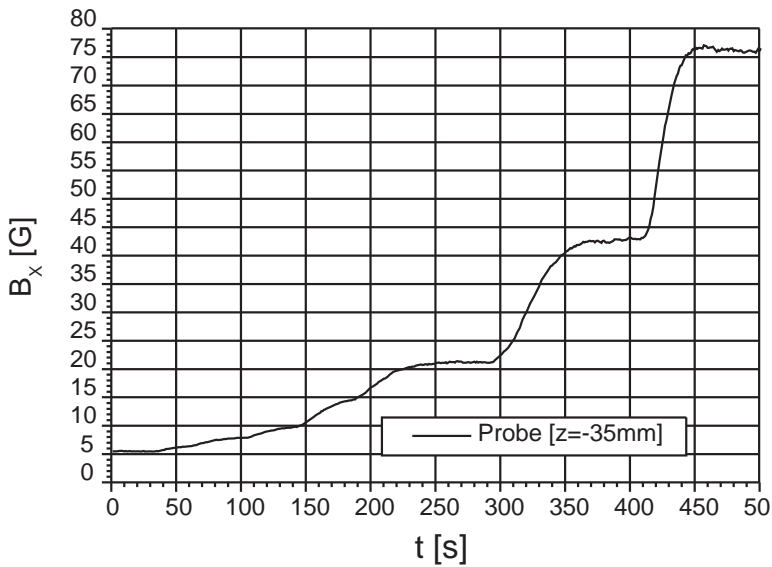}
\end{center}
\caption{Time recording of one component of the magnetic field in the Karlsruhe experiment. The amplitude of the magnetic field increases after each increasing step of the flow rate of liquid sodium. 
Small fluctuations are visible once the magnetic field has saturated at a constant mean value. (Figure from Stieglitz and M\"uller 2002).}
\label{karlsruhesignal}
\end{figure}

When the flow rates are large enough, a magnetic field is generated  
by dynamo action. The bifurcation is stationary and the magnetic  
field displays fluctuations caused by the small scale
turbulent velocity field (see figure \ref{karlsruhesignal}). This  
generation comes at a cost in the power necessary to drive the flow  
and the pressure drop increases.

\begin{figure}[!htb]
\vspace{0.0 in}
\begin{center}
\includegraphics[width=4 in]{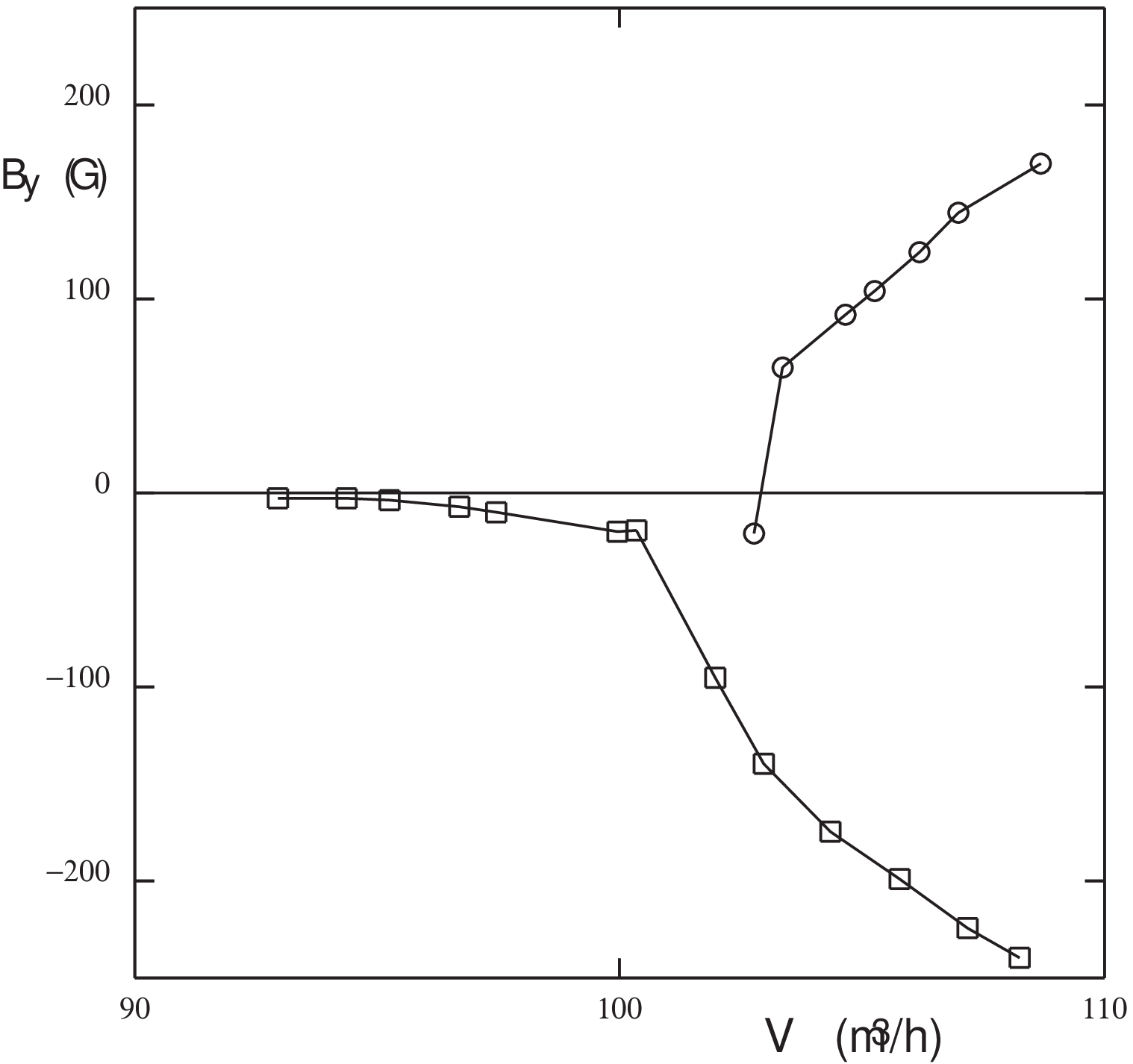}
\end{center}
\caption{The magnetic field amplitude increases above the critical  
flow rate
in the Karlsruhe experiment.  Another branch of self-generation can  
be reached only
by imposing an initial field.  This other branch is disconnected
from the main branch.  The imperfection of the bifurcation has been  
ascribed
to the Earth's field. (Figure from Stieglitz and M\"uller  
2002).}
\label{karlsruhebif}
\end{figure}

Due to the Earth's magnetic field, the bifurcation is imperfect but  
both branches of the bifurcation can be reached by applying an  
external magnetic field, as displayed in figure \ref{karlsruhebif}.  
Among others, the experimentalists performed careful studies of the  
dependence of the dynamo threshold  on the axial and helical flow  
rates and on the the electrical conductivity that can be varied by  
changing the temperature.  They also considered the effect of flow  
modulation on the dynamo threshold and studied the amplitude and the  
geometry of the magnetic field in the supercritical regime.

\subsection{The Riga experiment}

The experiment carried out by Gailitis et al. (2001) has been  
motivated by one of
the simplest examples of a homogeneous dynamo found by Ponomarenko (1973).
A conducting cylinder of radius $R$, embedded in an infinite static  
medium of the same
conductivity with which it is in perfect electrical contact, is in solid
body rotation at angular velocity $\Omega$,  and in translation along  
its
axis at speed $V$. In an unbounded domain,  this helical motion  
generates a
travelling wave magnetic field. This Hopf bifurcation occurs for a  
minimum critical magnetic
Reynolds number $R_{mc}=\mu_0 \sigma R \sqrt{(R \omega)^2+V^2} = 17.7$
for an optimum Rossby number $Ro=V/(R\Omega)=1.3$. We note that the  
maximum dynamo
capability of the flow ($R_{mc}$ minimum) is obtained when the  
azimuthal and axial velocities
are of the same order of magnitude ($Ro \sim 1$). This trend is often
observed with more complex flows for which the maximum dynamo capability
is obtained when the poloidal and toroidal flow components are  
comparable.

The experiment set up by the Riga group is sketched in figure \ref 
{karlsruheriga}b.
Their flow is driven by a single propellor, generating helical flow  
down a
central cylindrical cavity.  The return flow is in an annulus  
surrounding
this central flow.  The geometry of the apparatus as well as mean  
flow velocity profiles
have been optimized in order to decrease the dynamo threshold. In  
particular, it has been found that adding an outer cylindrical region  
with liquid sodium at rest significantly decreases $R_{mc}$.  This  
can be understood if the axial mean flow as well as the rotation rate  
of the azimuthal mean flow are nearly constant except in boundary  
layers close to the inner cylinder. Then, the induction equation  
being invariant under transformation to a rotating reference frame  
and under Galilean transformations, the presence of some electrical  
conductor at rest is essential as it is in the case of the  
Ponomarenko dynamo.
The three cylindrical chambers are separated by thin stainless steel  
walls, which
were wetted to allow currents to flow through them.

\begin{figure}[!htb]
\vspace{0.0 in}
\begin{center}
\includegraphics[width=5in]{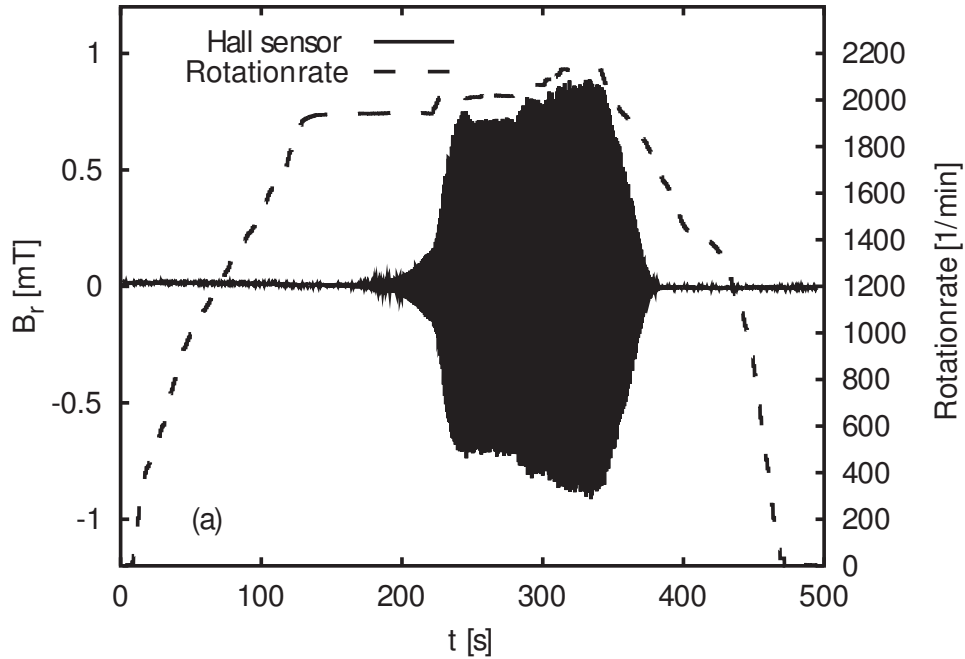}
\end{center}
\caption{Time recording of the magnetic field from the Riga  
experiment. The dashed line gives the value of the rotation rate. The amplitude of roughly sinusoidal oscillations (not visible with the resolution of the picture)  increases and then saturates when the rotation rate is increased above threshold (about $1850\, {\rm rpm}$). The amplitude of saturation increases when the rotation rate is increased further. The dynamo switches off when the rotation rate is decreased below threshold (Figure from Gailitis {\it et al.} 2001). }
\label{rigasignal}
\end{figure}

\begin{figure}[!htb]
\vspace{0.0 in}
\begin{center}
\includegraphics[width=5 in]{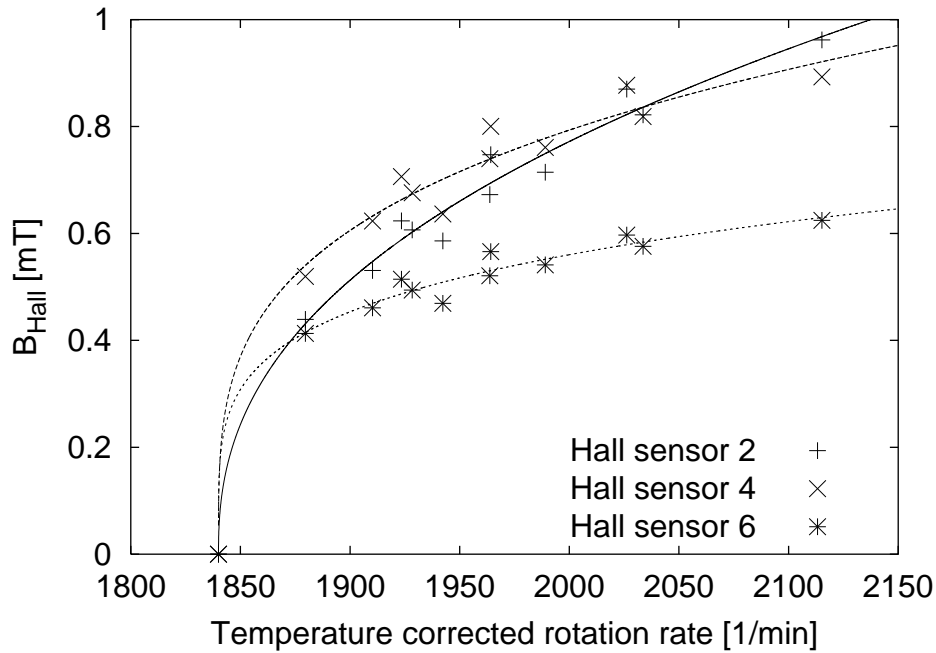}
\end{center}
\caption{Magnetic field amplitude as the rotation rate is raised
above the critical rotation rate. (Figure from Gailitis {\it et al.} 2003)}
\label{rigabif}
\end{figure}

Figure \ref{rigasignal} displays the growth and saturation of a time  
periodic magnetic field
at high enough rotation rate. The nature of the bifurcation as  
well as dynamo growth rates
have been found in good agreement with kinematic theory (Gailitis {\it et  
al.} 2002) that predicts a  Hopf bifurcation of convective nature at onset. Note that this bifurcation can be affected by the ramping time scale of the propellor rotation rate (Knobloch 2007). In addition,  the Riga group has made detailed  
observations of the magnetic field saturation value
and the power dissipation needed to drive the flow.  These  
measurements give
indications of the effect of Lorentz forces in the flow in order to  
reach the saturated state. It has been found that one effect of the  
Lorentz force is to drive the
liquid sodium in the outer cylinder in global rotation, thus  
decreasing the effective azimuthal velocity of the inner flow and  
therefore its dynamo capability. Dynamo generation does also  
correspond to an increase in the required mechanical power. However,  
a puzzling result
is displayed in figure \ref{rigabif}: the amplitude of the magnetic  
field for supercritical
rotation rates does not seem to show the universal $\sqrt{R_m - R_ 
{mc}}$ law.
In addition, the form of the law seems to depend on the location of  
the measurement
point. This is to some extent due to the absence of temperature  
control in the Riga experiment. Variations in temperature modify the  
fluid parameters (electrical conductivity, viscosity and density) and  
this should be taken into account by plotting the results in  
dimensionless form (Fauve and Lathrop 2005).

\subsection{Lessons from the Karlsruhe and Riga experiments}

Although there were no doubts about self-generation
of magnetic fields by Roberts' or Ponomarenko-type laminar flows, these
experiments have displayed several interesting features:
\begin{itemize}
\item the observed thresholds are in rather good agreement
with theoretical predictions (Busse {\it et al.} 1996, R\"adler {\it et al.} 1998,
Gailitis {\it et al.} 2002) made by considering only the laminar mean flow  
and neglecting the
small-scale turbulent fluctuations that are present in both experiments.
\item The nature of the dynamo bifurcation, stationary for the Karlsruhe
experiment or oscillatory (Hopf) in the Riga experiment, is also in
agreement with laminar models. \item On the contrary, the saturation  
level
of the magnetic field, due to the back reaction of the Lorentz force on
the flow, cannot be predicted with a laminar flow model and different
scaling laws exist in the supercritical dynamo regime depending on the
magnitude of the Reynolds number (P\'etr\'elis and Fauve 2001). \item
Although secondary instabilities generating large scale dynamics of the 
magnetic field (such as field reversals for instance) have not been observed 
in the Karlsruhe and Riga experiments, small scale turbulent fluctuations 
of the magnetic field  
are well developed. \end{itemize} These observations raise the following
questions: \begin{itemize}
\item What is the effect of turbulence, or of
the magnitude of the Reynolds number, on the dynamo threshold $R_{mc} 
$? Is
it possible to observe how $R_{mc}$ depends on $P_m$ for a dynamo generated
by a strongly turbulent flow (by changing $P_m$ in experiments 
with a given flow at different temperatures for instance)? 

\item What is the
mechanism responsible for magnetic field fluctuations in the vicinity of
the dynamo threshold: an on-off intermittency effect (Sweet {\it et al.}  
2001)
or chaotic advection of the mean magnetic field by the turbulent flow?
\item What is the mechanism for field reversals? Is it possible to  
observe
them in laboratory experiments? \end{itemize}

\section{Effect of turbulence on the dynamo threshold and saturation}

\subsection{Effect of small scale turbulent fluctuations on the  
dynamo threshold}

As said above, dynamo experiments involve high Reynolds number flows and
turbulent velocity fluctuations. Using the Reynolds decomposition, we  
write

\begin{equation}
\bf{V}(\bf{r}, t) = \overline {\bf{V} }(\bf{r}) + \tilde {\bf{v}} (\bf{r}, t) \,,
\label{reynolds}
\end{equation}
where $\overline {\bf{V} }(\bf{r})$ is the mean flow and $\tilde {\bf{v}} (\bf 
{r}, t)$ are
the turbulent fluctuations. The over-bar stands for a  
temporal average in experiments.
The induction equation then becomes
\begin{equation}
\frac{\partial{\bf {B}}}{\partial{t}} = \bf{\nabla} \times
(\overline {\bf{V} } \times \bf {B}) + \bf{\nabla} \times {(\bf 
{\tilde v}}
\times \bf {B}) + \nu_m \bf{\nabla}^2 \bf {B},
\label{indfluct}
\end{equation}
where $\nu_m = 1/(\mu_0 \sigma)$ is the magnetic diffusivity.
The Karlsruhe and Riga experiments have been designed by considering  
only
the mean flow $\overline {\bf{V} }(\bf{r})$ in order to predict the  
value of the
dynamo threshold. However, we observe that turbulent fluctuations 
$\tilde {\bf{v}}(\bf{r}, t)$
act as a random multiplicative forcing in the induction equation (\ref 
{indfluct}).
It is well known, both from simple theoretical models (Stratonovich    
1963, Graham and Schenzle 1985, L\"ucke  and Schank 1985)
and from experiments on different instability problems (Kabashima {\it et  
al.} 1979, Residori {\it et al.} 2001, Berthet {\it et al.} 2003, P\'etr\'elis  
and Auma\^{\i}tre 2003, P\'etr\'elis {\it et al.} 2005), that multiplicative  
noise generally shifts the bifurcation threshold.
In addition, it sometimes modifies the dynamics of the unstable modes  
in the vicinity of threshold. In particular,
multiplicative random  forcing may generate intermittent bursting in the
vicinity of instability onset (John {\it et al.} 1999, Berthet {\it et al.}  
2003, Auma\^\i tre and P\'etr\'elis 2003).
This type of behaviour has been also observed with deterministic chaos instead of noise. It has been understood  in the framework of  blowout  bifurcations in dynamical system theory (Platt {\it  et al.} 1993), and has been observed in a numerical simulation of the MHD equations without external noise (Sweet {\it et al.} 2001).  
Note however that in these simulations, $P_m$ is of order one, and the flow is chaotic at the dynamo threshold but not fully turbulent. We will use the same terminology ``on-off intermittency" both for deterministic dynamical systems and systems with a noisy bifurcation parameter.

The effect of velocity fluctuations has been considered a long time ago in order to explain reversals of the magnetic field of the Earth as a result of statistical fluctuations of cyclonic convective cells (Parker 1969). Similar models using a noisy $\alpha$-effect have been developed (Hoyng 1993). We will not study this type of problems here.

As said above, no significant shift in threshold with respect to
computations taking into account only the mean flow have been  
observed in the Karlsruhe and Riga experiments. Bursting phenomena in the vicinity of  threshold have not been reported either despite the existence of velocity fluctuations. 

An explanation for
the small shift in threshold has been given by P\'etr\'elis (2002) and  
Fauve and P\'etr\'elis (2003) as follows: in the limit of small fluctuations, one can calculate the  
threshold shift using a perturbation expansion.
Let $\overline {\bf{V} }$ be the average velocity field at onset, and $\mathbf B$ the neutral mode of the instability. Let ${\bf V}^{(0)}$ be the flow leading to the neutral mode $\mathbf B^{(0)}$ when there is no velocity fluctuations.  Our aim is to find how the dynamo threshold of the velocity field ${\bf V}^{(0)}$ is modified in the presence of small turbulent fluctuations. We write $\tilde {\bf{v}} = \delta \bf{v}$ where $\delta$ is a small parameter that measures the intensity of the turbulent fluctuations so that the amplitude of $\bf{v}$ is of order one. The neutral mode is likely to be slightly modified by the fluctuations as well as the dynamo threshold. We expand $\mathbf B$ and $\overline {\bf{V} }$ in powers of $\delta$
\begin{equation}
{\bfB}={\bfB}^{(0)}+\delta\,{\bf B}^{(1)}+\delta^2\,{\bf B}^{(2)}+.
..\,,\nonumber\\
\overline {\bf{V} } = {\bf V^{(0)}} (1+ c_1 \delta\,+ c_2 \delta^2\,  
+... )\,,\label{devbruit}
\end{equation}
$\mathbf B^{(i)}$ are the corrections at order $i$ to the neutral mode due to the presence of the turbulent fluctuations. $c_i$ are constants that express the shift in the dynamo threshold caused by turbulence. 
We emphasise that we study the modification of the dynamo threshold of a mean flow with prescribed geometry due to the presence of fluctuations. When one inputs these expressions in equation (\ref{indfluct}), the zeroth order part can be written
\begin{equation}
L\, \bf{B}^{(0)}=\frac{\partial \bf{B}^{(0)}}{\partial
t}-\,  \bf{\nabla} \times \left(V^{(0)} \times \bf{B}^{(0)}\right)  -  
\nu_m\, \bf{\nabla}^2 \bf{B}^{(0)}=0\,.
\label{eqBbruitordre0}
\end{equation}
$L$ being the linear operator in the l.h.s. member. 
This is the laminar dynamo problem. By hypothesis, the instability  
onset is the one  without turbulent perturbation. At next order in $\delta$ we get
\begin{equation}
L\,{\bf B^{(1)}}= c_1\, \bf{\nabla} \times \left(V^{(0)} \times \bf{B}^ 
{(0)}\right) + \bf{\nabla} \times \left(  \bf{v} \times \bf{B}^{(0)} 
\right).
\label{eqBbruitordre1}
\end{equation}
We now introduce a scalar product $\langle f|g\rangle$ and calculate $L^+$ the adjoint of $L$.
As $LB^{(0)}=0$, $L^+$ also has  a nonempty kernel. Let ${\bf C}$ be in this kernel. Then
\begin{equation}
\langle{\bf  C}|L{\bf B^{(1)}}\rangle= \langle L^+{\bf C}|{\bf B^{(1)}}\rangle=0
\end{equation}
and this solvability condition gives the first order correction in the threshold

\begin{equation}
c_1=-\frac{\sca{\bf{C}}{\bf{\nabla} \times \left( \bf{v} \times \bf{B} 
^{(0)}\right)}}{\sca{\bf{C}}{\bf{\nabla} \times
\left(V^{(0)} \times \bf{B}^{(0)}\right) }}\,,
\label{premierecorrectionRm}
\end{equation}

We use a scalar product in which the average over the realisations of
the perturbation is made. In that case, the average over the
realisations of $\sca{\bf{C}}{\bf{\nabla} \times \left(  \bf{v}
\times \bf{B}^{(0)}\right)}$ is proportional to the average of
$\bf{v}$, the value of which is zero by hypothesis. Thus, the dynamo
threshold is unchanged up to first order in $\delta$, $c_1=0$.
This result is obvious in many simple cases. For instance if $\tilde {\bf{v}}$ is sinusoidal in time, the threshold shift cannot depend on the phase which implies that it is invariant if $\tilde {\bf{v}} \rightarrow - \tilde {\bf{v}}$. This is also true if $\tilde {\bf{v}}$ is a random noise with equal probabilities for the realisations $\tilde {\bf{v}}$ and $-\tilde {\bf{v}}$. Note     however that simple symmetry arguments do not apply for asymmetric fluctuations about the origin although the threshold shift vanishes to leading order if the fluctuations have zero mean.  

To calculate the next order correction, we write equation (\ref 
{indfluct}) at order two in $\delta$ and get
\begin{equation}
L\,{\bf B}^{(2)}= c_2\,\bf{\nabla} \times \left(V^{(0)} \times \bf{B}^ 
{(0)}\right) + \bf{\nabla} \times \left( \bf{v} \times \bf{B}^{(1)} 
\right).
\label{eqBbruitordre2}
\end{equation}
We then get the second order correction
\begin{equation}
c_2 =-\frac{\sca{\bf{C}}{\bf{\nabla} \times \left(\bf{v} \times \bf{B} 
^{(1)}\right)}} {\sca{\bf{C}}{\bf{\nabla} \times
\left(V^{(0)} \times \bf{B}^{(0)}\right)}}\,,
\label{deuxiemecorrectionRm}
\end{equation}
where $\bf{B}^{(1)}$  is solution of
\begin{equation}
L\bf{B}^{(1)}=\bf{\nabla} \times \left( \bf{v} \times \bf{B}^{(0)} 
\right).
\label{eqB1bruite}
\end{equation}
Here, there is no simple reason for the correction to be zero. Its  
computation requires the resolution of equation (\ref{eqB1bruite}).
In some simple cases, an analytical expression for $c_2$ can be  
calculated and both signs can be found, thus showing that  
fluctuations can in general increase or decrease the dynamo threshold  
(P\'etr\'elis and Fauve 2006).

We have thus obtained that when the amplitude $\delta$ of turbulent   
fluctuations is small, the modification of the dynamo threshold  is  
at least quadratic in $\delta$.  Consequently, we can understand why  
the thresholds measured in the Karlsruhe and Riga experiments are  
very close to the
predictions using the mean flow $\overline {\bf{V} }$ and thus  
ignoring turbulent fluctuations (the order of magnitude of the level  
of turbulent fluctuations related to the mean flow is certainly less  
than $10\%$ in these experiments). The problem is more complex for  
experiments with unconstrained flows
for which large scale fluctuations can be of the same order as the  
mean flow. We will consider it in the section about ``turbulent  
dynamos".

\subsection{Scaling laws for magnetic energy density in the vicinity  
of threshold}

In order to describe the saturation of the magnetic field above the dynamo  threshold, we need to take into account its back reaction 
on the velocity field. We thus have to solve the induction
and Navier-Stokes equations that we restrict to incompressible flows  
(${\bf \nabla}\cdot {\bf V} = 0$),
\begin{equation}
\frac{\partial {\bf B}}{ \partial t} =  \bf{\nabla} \times ({\bf V}  
\times {\bf B}) + \frac{1}{\mu_0 \sigma} \bf{\nabla}^2 {\bf B}, \label 
{ind}
\end{equation}
\begin{equation} \frac{\partial {\bf V}}{ \partial t} + ({\bf V}  
\cdot \bf{\nabla}) {\bf V} = -  {\bf \nabla} \left(\frac{p }{ \rho} +  
\frac{B^2 }{ 2\mu_0}\right) + \nu \bf{\nabla}^2 {\bf V}
+ \frac{1}{ \mu_0 \rho} ({\bf B} \cdot \bf{\nabla}) {\bf B}. \label{ns}
\end{equation}
The flow is created, either by moving solid boundaries or by a body  
force added to the Navier-Stokes equation. We do not consider global  
rotation ($\Omega = 0$).
We have to develop equations (\ref{ind}, \ref{ns}) close to the  
dynamo threshold in order to derive an amplitude equation for the  
growing magnetic field. If the
dynamo bifurcation is found supercritical, this allows the calculation of   
the saturated
mean magnetic energy density $\langle  B ^2 \rangle / 2\mu_0$, 
where $\langle \cdot \rangle$ stands for average in both space and  
time in this section.

Thus, even in the simplest configuration, the problem involves three  
dimensionless
parameters. One can choose,  the magnetic Reynolds number, $R_m$,   
the magnetic Prandtl number, $P_m$,
and the ratio of the mean magnetic to kinetic energy density,   
leading to the following form of law
\begin{equation}
\frac{\langle B^2 \rangle}{\mu_0} = \rho\langle V^2 \rangle \, f(R_m,  
P_m).
\label{dim}
\end{equation}
In general, the analytic determination of  $f$ using weakly nonlinear  
perturbation theory
in the vicinity of the dynamo threshold is tractable only in the  
unrealistic case $P_m \gg 1$ such that the dynamo bifurcates from a  
laminar flow ($Re \ll 1$). For $P_m \ll 1$, a lot of hydrodynamic  
bifurcations occur first and the flow becomes turbulent before the  
dynamo threshold that occurs for $Re \gg 1$. We will briefly recall  
the scaling laws that are expected in both limits (for more details,  
see P\'etr\'elis and Fauve 2001).

\subsubsection{Saturation in the low Re limit}

At small $Re$, the Lorentz force should be balanced by the viscous   
force as it is larger than the inertial one in (\ref{ns}). When the  
magnetic field bifurcates above a critical velocity amplitude $V_c$,  
it thus generates a velocity modification of order $\delta V$ given by
\begin{equation}
\nu \frac{\delta V}{L^2} \propto \frac{B^2}{\rho \mu_0 L}.
\end{equation}
If this bifurcation is supercritical, we expect saturation for $ 
\delta V$ of the order of the distance to criticality $V-V_c$. In the  
bifurcation analysis, this balance results from the solvability  
condition. We thus obtain
\begin{equation}
\langle B^2 \rangle   \propto \frac{\rho \nu}{\sigma L^2} \, (R_m - R_ 
{mc}).
\label{laminarscaling}
\end{equation}

Analytic calculations in the small $Re$ limit have been performed  
both for the Roberts flow (Gilbert and Sulem  1990, Busse and  
Tilgner 2001) and Ponomarenko type flows (Nu\~nez {\it et al.} 2001). In  
the case of the Roberts flow for which there are two spatial scales,  
the flow periodicity $l$ and the size $L$ of the full flow volume,  
the equation for $\langle B^2 \rangle$ in the limit $L \gg l$ takes the form
\begin{equation}
\langle B^2  \rangle  \propto \frac{\rho \nu}{\sigma l^2} \, (R_m - R_ 
{mc}),
\label{robertsscaling}
\end{equation}
with $R_m = \mu_0 \sigma \sqrt{UVLl}$ where $U$ and $V$ are  
respectively the axial and azimuthal typical velocities. The  
occurrence of $l$ instead of $L$ results from the balance between the  
Stokes and the Lorentz forces that both imply the small scale. The  
largest contribution to the current density is indeed the one  
associated to the small scale magnetic field in the limit of scale  
separation.

More surprisingly, the basic property of this ``laminar scaling",  
i.e.,  $B^2$ proportional to $\nu$, or $B^2/\mu_0 \rho V^2 \propto 1/Re$ in dimensionless form, subsists in the case of some rotating flows  (Childress and Soward 1972, Soward 1974).  
Equation (\ref{laminarscaling}) is known as the ``weak field  
scaling" of the Earth dynamo (Roberts 1988). The same property has been also found for a smooth helical flow in the limit $Re \gg R_m \gg 1$ (Bassom and Gilbert 1997).

\subsubsection{Saturation in the large Re limit}

At large $Re$, balancing the Lorentz force with the inertial one gives
\begin{equation}
\frac{V_c \delta V}{L} \propto \frac{B^2}{\rho \mu_0 L}.
\end{equation}
Then, using $\delta V \propto V-V_c$, we obtain the large $Re$ scaling
\begin{equation}
\langle B^2 \rangle  \propto \frac{\rho}{\mu_0 (\sigma L)^2} \, R_ 
{mc} (R_m - R_{mc}).
\label{turbulentscaling}
\end{equation}
The dimensionless group ${\langle B^2 \rangle \mu_0 (\sigma L)^2/  
\rho}$ is known as the Lundquist number.

The form of (\ref{turbulentscaling}) can be easily found by  
dimensional arguments. For $P_m \ll 1$, the Ohmic dissipative scale,  
$l_{\sigma} = L R_m^{-3/4}$, is much larger than the Kolmogorov  
scale, $l_K = L Re^{-3/4}$. Thus,  the magnetic field grows at scales  
much larger than $l_K $ and does not depend on kinematic viscosity.  
Then, we can discard the dependence of $f$ on $P_m$ in (\ref{dim}).  
In the vicinity of dynamo threshold, $V$ is not a free parameter but  
is such that $\mu_0 \sigma V L \approx R_{mc}$. We obtain
\begin{equation}
\langle B^2 \rangle  \propto \frac{\rho}{\mu_0 (\sigma L)^2} \, f(R_m).
\label{largerescaling}
\end{equation}

The scalings for large and small $Re$ differ by a factor $R_{mc} / P_m  
\approx 10^6$ for experiments using liquid sodium ($P_m \approx 10^ 
{-5}$).
It may be instructive to replace $\nu$ by the turbulent viscosity, $ 
\nu_{\rm T} \propto VL$ (respectively $\nu_{\rm T} \propto Vl$ in the  
case of scale separation)
in the laminar scaling (\ref{laminarscaling}). Using $V \approx R_ 
{mc} / \mu_0 \sigma L$, we have
\begin{equation}
\langle B^2 \rangle \propto \frac{\rho \nu_{\rm T}}{ \sigma L^2} \;  
(R_m - R_{mc})
\propto \frac{\rho }{ \mu_0 (\sigma L)^2} \; (R_m - R_{mc}) \,.\label 
{viscturb}
\end{equation}
We thus recover the turbulent scaling. However, the above analysis  
does not require any assumption about the expression of the  
turbulent viscosity and is thus clearer.

\subsubsection{Magnetic energy density of the Karlsruhe and Riga dynamos}

\begin{figure}[!htb]
\begin{center}
\includegraphics[width=.75\textwidth]{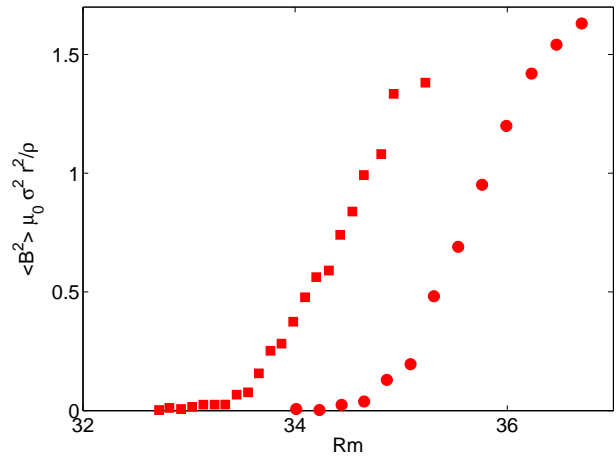}
\caption{Dimensionless magnetic energy density $\langle B^2 \rangle \mu_0 \sigma^2 r^2/\rho$  
as a function of the magnetic Reynolds number $R_m$ for the Karlsruhe  
experiment. The different symbols are associated to different ratios  
of azimuthal to axial velocity. $R_m$ is varied through change in the  
temperature leading to a variation of the electrical conductivity  
(data from M\"uller {\it et al.} 2004).}
\label{figkarlsruhe}
\end{center}
\end{figure}

The Lundquist number, $\langle B^2 \rangle \mu_0 (\sigma r)^2 / \rho$  
where $r$ is the radius of each cylinder of the Karlsruhe
experiment, is plotted as a function of $R_m$ in figure \ref 
{figkarlsruhe} for two different ratios of the axial to azimuthal  
flows. The large scale magnetic field being the dominant component  
and slowly varying in space, its local value thus gives a reasonable  
estimate of the mean energy density.
The choice of the small scale $r$ instead of the full size of the  
flow in the Lundquist number results from equation (\ref 
{robertsscaling}).
$R_m$ has been defined using the geometrical mean of the axial and  
helical velocities as velocity scale but the height of the  
cylindrical volume $H=0.7\, {\rm m}$ has been taken as length scale. The later  
choice is somewhat arbitrary but allows a simple comparison with the  
other fluid dynamos.

Figure \ref{figkarlsruhe} shows that the dynamo threshold depends on  
the axial to azimuthal flow ratio although its leading order effect  
has been taken into account in the
definition of the velocity scale in $R_m$. This may result from the  
expulsion of the transverse magnetic field by the azimuthal flow. 
Another possible explanation comes from the geometry of the helical flow.  Indeed, part of the helical flow contributes to the axial flow. Taking this component into account in the definition of $R_m$ could decrease the difference between critical magnetic Reynolds numbers.  
On the contrary, the slope of the bifurcation curve does not seem to be strongly affected by the axial to azimuthal flow ratio.

\begin{figure}[!htb]
\begin{center}
\includegraphics[width=.75\textwidth]{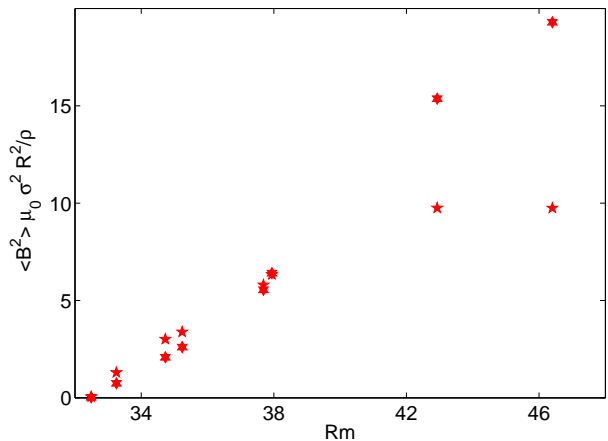}
\caption{Dimensionless magnetic energy density $\langle B^2 \rangle \mu_0 \sigma^2 R^2/\rho$  as a function of the magnetic Reynolds number $R_m$ for the Riga experiment. The different values for same $R_m$ are associated to different probe  positions (data courtesy of Franck Stefani).}
\label{figriga}
\end{center}
\end{figure}

Figure \ref{figriga} displays the dynamo bifurcation using the same  
parameters for the Riga experiment. The Lundquist number as well as  
$R_m$
are defined using the radius $R=0.125\, {\rm m}$ of the inner cylinder. The  
two curves are related to probes at different locations. We observe  
that the imperfection ascribed to the Earth's magnetic field in the  
Karlsruhe experiment does not exist in the Riga experiment. A  
constant external field is not resonant with the neutral mode that is  
a travelling wave and thus does not lead to any imperfection.  Other  
measurements (not shown here) made at different spatial locations  
(see figure \ref{rigabif}) have displayed smaller values of the  
magnetic field that is concentrated in the vicinity of the shear for  
a Ponomarenko-type dynamo.  Multiple point measurements and averaging  
thus should be made in order to have a better evaluation of the mean  
magnetic energy density in the flow volume. Fortunately, the  
difference between the predictions (\ref{laminarscaling}) and (\ref 
{turbulentscaling}) is so large that rough order of magnitude  
estimates of $\langle B^2 \rangle$ are enough.
Taking into account the qualitative nature of our analysis in section  
3.2, we conclude that the large $Re$ scaling is in agreement with the  
experimental observations whereas the ``laminar scaling" predicts a  
field that is orders of magnitude too small.
We thus note that Karlsruhe and Riga experiments display an  
interesting feature: turbulent fluctuations can be neglected when  
computing the dynamo threshold whereas the high value of $Re$ has a  
very strong effect on the amplitude of the saturated magnetic energy  
density above the dynamo threshold.

\section{Laminar versus turbulent dynamos}

In a fully developed turbulent flow in a compact domain, i.e.,  
without any mean flow prescribed by strong geometrical  
constraints, large scale fluctuations are generally of similar order  
of magnitude as the mean flow. Then, we cannot a priori expect that  
the dynamo computed as if it were generated by $\overline {\bf{V} }(\bf{r}) 
$ alone can describe the observations as in the limit of small
fluctuations considered in section 3.1. We thus propose the following  
definition: we call ``laminar" a fluid dynamo which generates a  
magnetic field that displays the same characteristics as the one  
computed as if the mean flow  $\overline {\bf{V} }(\bf{r})$ were  
acting alone. We mean by characteristics, the geometry of the mean  
field lines, the nature of the bifurcation (stationary, Hopf, etc)  
and the related dynamics of the large scale field, the approximate  
value of $R_{mc}$, etc. Otherwise, the dynamo is called turbulent,  
which means that turbulent velocity  fluctuations $\tilde {\bf{v}}(\bf 
{r}, t)$ qualitatively modify the nature of the bifurcation obtained  
with $\overline {\bf{V} }(\bf{r})$ alone. This of course does not  
mean that $\overline {\bf{V} }(\bf{r})$ has no effect on the dynamo  
process. This would remain true even if $\overline {\bf{V} }(\bf{r})$ alone has no dynamo  
capability.

\subsection{Turbulent fluctuations and the shift of the dynamo threshold}

It has been believed that studying the dynamo generated by the mean flow can
often provide a leading order description of a fluid dynamo even in the
presence of large fluctuations. As just said, this is far from being
obvious, but in that case, a natural question is related to the effect of
turbulent fluctuations on threshold as considered in section 3.1 in the
limit of small fluctuations. There exists no general answer to this
question. Even in the limit of scale separation, it is known that the  role
of turbulent fluctuations  may be twofold: on one hand, they decrease the
effective electrical conductivity and thus inhibit the dynamo action
generated by $\overline {\bf{V} }(\bf{r})$ by increasing Joule
dissipation; on the other hand, they may generate a large scale magnetic
field through the ``$\alpha$-effect" (Krause and R\"adler 1980, Moffatt
1978) or higher order similar effects even if $\overline {\bf{V} }(\bf{r})
= 0$.  However, most natural flows do not involve any clear-cut scale
separation, and it has been recently shown in several numerical simulations
that the ``$\alpha$-effect" may be very weak even for flow configurations
involving large helicity where it could be expected to drive an efficient
large scale magnetic field (Cattaneo and Hughes 2006, Hughes and Cattaneo
2007).

The effect of turbulent fluctuations on the dynamo threshold has been also
studied with direct numerical simulation of the Taylor-Green flow, but with different outcomes: the observed threshold has been
found to increase in the presence of large scale fluctuations (Laval {\it et al.}
2006) whereas it has been observed to correspond  to some threshold
corresponding to the mean flow alone  (Ponty {\it et
al.} 2005, 2007). It should be emphasised that these
direct simulations, as well as the previous ones about the
``$\alpha$-effect", are not performed with small enough values of $P_m$ so that one must be careful when using them to understand laboratory
experiments.

In order to try to clarify this issue, it is instructive to consider the
evolution equation of
the magnetic energy
\begin{equation}
\frac{d}{ dt} \int \frac{B^2}{2 \mu_0} = \int  ({\bf V} \times {\bf B})
\cdot {\bf j} \, d^3 x - \int  \frac{j^2}{\sigma} \, d^3 x,
\label{energy1}
\end{equation}
where ${\bf j}$ is the current density ($\mu_0 {\bf j} \approx \bf{\nabla} \times \bf{B}$ in the MHD approximation).
At the dynamo threshold, the first term on the right hand side, the
amplification term, should be equal to the second one that corresponds to
Ohmic dissipation. If we assume that this is achieved in the absence of
fluctuations, we observe that turbulent
fluctuations can increase the dynamo threshold through two different types
of mechanism: the most obvious one results from an increase of Ohmic losses
due to the generation of magnetic field at small scales by advection of
field lines by turbulent eddies. The second one is a loss of the efficiency
of field amplification. The amplification term depends on the relative
orientation of ${\bf V}$, ${\bf B}$ and ${\bf j}$, thus fluctuations in
direction of these vectors can decrease it. We briefly discuss these two
phenomena.

The increase of Ohmic dissipation due to turbulence primarily depends on the
value of the magnetic Reynolds number related to the corresponding velocity
fluctuations, $\mu_0 \sigma l v_{rms}$ where $v_{rms}$ is the $rms$ value of velocity fluctuations
$\tilde {\bf{v}}$ and $l$ is their integral scale. If $\mu_0 \sigma l v_{rms} \approx 1$ or smaller, the
fluctuations of the magnetic field due to turbulence have a spectrum of the
form $\vert \hat B \vert^2 \propto (\mu_0 \sigma B_0)^2 k^{-2} \vert \hat v \vert^2 
\propto (\mu_0 \sigma B_0)^2 \epsilon^{2/3} k^{-11/3}$ where $\bf{B}_0$ is the large scale
magnetic field and $\epsilon = V^3/L$ is the energy flux per unit mass
(Golitsyn 1960, Moffatt 1961). This
spectrum being steeper than $k^{-3}$, currents related to the large scale
magnetic field are dominant and Ohmic dissipation is hardly increased by
turbulence. The dynamo threshold is only slightly shifted as explained in
section 3
and observed in Karlsruhe and Riga experiments for which $\mu_0 \sigma l
v_{rms}$ is of order one or less.

The situation strongly differs in flows without geometrical constraints for
which Kolmogorov-type  turbulent fluctuations exist up to the size $L$ of the
fluid container and have a $rms$ value of the same order as the mean flow.
The magnetic Reynolds number of turbulent fluctuations is thus of the same
order as the one related to the mean flow, $R_m$, and $R_m > 10$ at
threshold. Then there exists an inertial magnetic range, $l_{\sigma} = L
R_m^{-3/4} < l < L$, in which $\vert \hat B \vert^2$ follows a different
scaling law.  Unfortunately, no theoretical prediction for $\vert \hat B
\vert^2$ is available in the vicinity of the dynamo threshold. Note that predictions 
for spectra of magnetohydrodynamic turbulence
such as, $\vert \hat B \vert^2 \propto (\rho \mu_0)^{3/4} (\epsilon
B_0)^{1/2} k^{-3/2}$ (Iroshnikov 1964,  Kraichnan 1965) followed by
many other models based on Alfven wave turbulence (for a review, see Verma 
2004), require an applied field value $B_0$ larger than the one achieved in
the vicinity of the dynamo threshold. Similarly, the Kolmogorov-type
spectrum $\vert \hat B \vert^2 \propto \rho \mu_0 \epsilon^{2/3} k^{-5/3}$, 
recently proposed for
small $P_m$ turbulent dynamos at large $R_m$ (Fauve and P\'etr\'elis 2007),
is not expected close to threshold.

Dimensional analysis lead to
\begin{equation}
\vert \hat B \vert^2 \propto \rho\mu_0\langle V^2 \rangle k^{-1} f
(R_m,kl_\sigma) \, ,
\end{equation}
where $f$ is an arbitrary function.
In the vicinity of the dynamo threshold it is expected to become
\begin{equation}
\vert \hat B \vert^2 \propto \rho\mu_0\langle V^2 \rangle k^{-1} (R_m-
R_{mc}) \, g(kl_\sigma) \, ,
\end{equation}
where $g$ is another arbitrary function. Let us assume the following form of the spectrum of the
magnetic fluctuations:
\begin{align}
\vert \hat B \vert^2 &= A (k/k_\sigma)^\alpha && \textrm{ if  }k/k_\sigma<1\cr
&= A (k/k_\sigma)^{-11/3} && \textrm { if  }k/k_\sigma>1,
\end{align}
where $\alpha$ is an exponent that depends on the regime under consideration,
$k_\sigma=2\pi/l_\sigma$, and $A$ is obtained by $\int  \vert \hat B \vert^2
dk =\langle B^2\rangle$. For $\alpha > -1$, one gets $A \propto \langle B^2\rangle
LRm^{-3/4}$ and for $\alpha<-1$, $A \propto \langle B^2\rangle LRm^{3\alpha/4}$.

Then using
\begin{equation}
\frac{1}{\sigma} \langle {j}^2 \rangle  = \frac{1}{\sigma} \int \vert \hat j
\vert^2 \, {\rm d}k \propto  \frac{1}{\mu_0^2 \sigma} \int k^2 \vert \hat B
\vert^2 \, {\rm d}k\, ,
\label{joule}
\end{equation}
we estimate Ohmic dissipation

\begin{align}
\frac{1}{\sigma} \langle {j}^2 \rangle & \propto  \frac{\langle B ^2\rangle}{\mu_0^2\sigma L^2} Rm^{3/2} \phantom{aaaaaa} \textrm{ if
}\alpha >-1,\cr
& \propto  \frac{\langle B ^2\rangle}{\mu_0^2\sigma L^2} \frac{Rm^{3/2}}{{\rm Log}
(Rm)}\phantom{aaa,}\textrm{ if  }\alpha =-1,\cr
& \propto \frac{\langle B ^2\rangle}{\mu_0^2\sigma L^2} Rm^{(9+3\alpha)/4}\phantom{aaa}\textrm{if  }-3<\alpha <-1,\cr
&\propto \frac{\langle B ^2\rangle}{\mu_0^2\sigma L^2}\phantom{aaaaaaaaaaaa}\textrm{ if  }\alpha <-3.
\end{align}

The next step is to relate $\langle B^2\rangle$ to $\langle B\rangle$.
Following Moffatt (1961), one can assume that $\langle B\rangle$ imposes the
value of the spectrum at $k_L$ so that  $\langle B\rangle^2\approx
A(k_L/k_\sigma)^\alpha k_L$. Then $\langle B^2\rangle \propto \langle B\rangle^2
Rm^{(9+3\alpha)/4}$ if $\alpha>-1$ and $\langle B^2\rangle \propto \langle
B\rangle^2$ otherwise. One gets

\begin{align}
\frac{1}{\sigma} \langle {j}^2 \rangle & \propto \frac{\langle B\rangle^2}{\mu_0^2\sigma L^2} Rm^{(9+3\alpha)/4}\phantom{aaa} \textrm{
if  }\alpha > -3,\cr
&\propto \frac{\langle B\rangle^2}{\mu_0^2\sigma L^2}\phantom{aaaaaaaaaaaa,}\textrm{ if  }\alpha <-3,
\end{align}

$\langle B\rangle^2 / (\mu_0^2\sigma L^2)$ being the dissipation in the
laminar regime.
We thus obtain that the effect of turbulent fluctuations is to increase
Ohmic dissipation by a factor $R_m^{5/2}$ if $\alpha = 1/3$ according to  Moffatt (1961) or
by a factor $R_m^{3/2}$ if $\alpha = -1$ according to Ruzmaikin and Shukurov (1982).
To sum up, any spectrum less steep than $k^{-3}$ in the interval $[k_L, k_{\sigma}]$ leads to an enhancement of Ohmic dissipation.

Fluctuations can also increase the dynamo threshold in the absence of any
turbulent cascade. A simple example has been provided by P\'etr\'elis and
Fauve (2006) by considering phase fluctuations of the G. O. Roberts flow
\begin{equation}
{\bf V}=\begin{pmatrix}
V\left (\cos{(k y+\phi)}-\cos{(k z+\psi
)}\right)\\
U\, \sin{(k z+\psi)}\\
U\,\sin{(k y+\phi)}
\end{pmatrix}
\label{vphase}
\end{equation}
where $\psi$ and $\phi$ are two functions that depend on time only,
or
\begin{equation}
{\bf V}=\begin{pmatrix}
V\left (\cos{(k y+\phi)}-\cos{(k z+\psi)}\right)\\
U\, \sin{(k z+\psi)}-(V/k) \partial_x\phi \cos{(k y+\phi)}\\
U\,\sin{(k y+\phi)}+(V/k) \partial_x\psi \cos{(k z+\psi)}\end{pmatrix}
\label{vphasex}
\end{equation}
where $\phi$ and $\psi$ depend on $x$ only.
Both flows involve phase fluctuations, i.e., motions of the eddies. In the
first case, this amounts to switch the origin of the flow in time. In the
second case, spatial fluctuations of the eddies along the $x$-axis are
considered, the additional terms in the second and third components of the
velocity field are just ensuring the incompressibility of the flow. In the limit
of scale separation, and assuming that the phase gradients in time (respectively
in space) are small enough, it has been shown that these large scale
fluctuations always increase the dynamo threshold of the G. O. Roberts flow.
The above examples are simple enough to compute how fluctuations increase
the threshold of the neutral mode generated by the mean flow alone. They also show
that even if this threshold is increased,  different modes that cannot be
amplified by the mean flow,  can be generated by the time dependent flow. It
has been indeed shown that when $\psi$ and $\phi$ are sinusoidal functions of time
in the first flow given above, fast dynamo modes are generated (Galloway and
Proctor 1992).

In conclusion, no general statement about the effect of fluctuations on the
dynamo threshold can be made. When fluctuations occur at small scale and are of small amplitude such that their magnetic Reynolds number is small, their
effect on the neutral mode generated by the mean flow alone is small and can
be computed perturbatively. The shift in threshold occur at second (or
higher) order. On the contrary, for Kolmogorov type turbulence with large
scale fluctuations of the same order as the mean flow, it is expected that
Ohmic dissipation is increased and the efficiency of the amplification
mechanism of the neutral mode generated by the mean flow alone can be
decreased. However, other modes may be selectively amplified due to the
presence of fluctuations. These observations question the validity of dynamo
models based only on the mean flow, thus neglecting the effect of large
scale turbulent fluctuations. Experimental studies of the transport of a
localized magnetic field by a turbulent flow have also shed light on the
effect of fluctuations (Volk {\it et al.} 2006b). In particular, they show a loss
of the magnetic field orientation in the transport process that may lower
the efficiency of the field amplification as stated before.

\subsection{On-off intermittency}

In addition to a shift of threshold, the nature of the bifurcation  
can be modified in the presence of fluctuations. The simplest example  
is provided by the phenomenon of on-off intermittency.
On-off intermittent behaviour is a common feature for bifurcating  
systems subjected to multiplicative noise. Consider a system
close to an instability threshold. Multiplicative noise induces fluctuations of the
instantaneous departure from onset. If the fluctuations are large  
compared to the mean departure from onset, the average
over a time interval $T$ of the instantaneous departure from onset can be  
negative even for long duration $T$. Then the
amplitude of the unstable mode tends to zero (off-phase) before the  
average departure from onset turns back to a positive value and the  
unstable mode
reaches amplitudes where nonlinearities saturate its growth (on-phase). 
This is a simple mechanism that leads to the random  
succession of phases where the unstable system
is either close to its formerly stable solution or reaches values  
controlled by non-linearities.

Up to now this behaviour has not been displayed by experimental  
dynamos. Two key elements have been identified that can limit the  
ability for a dynamo to exhibit an on-off intermittent magnetic field.

The fluctuations that drive the off-phases are long time  
fluctuations. For simple dynamical systems, it  has been shown that  
the on-off intermittent behaviour is controlled
by the zero frequency component of the spectrum of the multiplicative  
noise (Auma\^ \i tre {\it et al.} 2005, Auma\^ \i tre {\it et al.} 2006).  
Indeed, on-off intermittency takes place when the ratio between the  
departure from
onset and the zero frequency component of the noise is small.
It seems reasonable to assume that for dynamos, fluctuations of the  
velocity field at very small frequencies are required. It is probable  
that the currently working dynamos
do not have enough low frequency fluctuations to generate an on-off  
intermittent magnetic field.

Another effect that can play a role is the imperfectness of the  
bifurcation. It is important for the mechanism of on-off  
intermittency that when the system evolves towards
the off-phase,  the effect of the noise term vanishes. It the  
bifurcation is imperfect, the deterministic solution does not tend to  
a zero magnetic field solution but to the imperfect branch.
Then the effect of multiplicative noise does not vanish. This  
generates fluctuations that can destroy the intermittent regime (P\'etr\'elis and Auma\^ \i tre 2006).
Note that the effect is  the same as for additive noise that also  
leads to a disappearance of on-off intermittency (Platt {\it  et al.}  
1994). In the dynamo context, even though the fluctuations
remain multiplicative, the imperfectness of the bifurcation can be  
the source of the fluctuations that prevent the observation of 
on-off intermittency.

A source of imperfectness for experimental dynamos is the ambient  
magnetic field of the Earth and of the possibly magnetised parts of  
the experiment, as for instance the
propellors in the VKS experiment. It might be helpful to screen  
these ambient fields in order to observe on-off intermittent dynamos.

To sum up, in order to favour on-off intermittency above the dynamo  
onset, it is important to lower the source of imperfectness of the  
bifurcation, for instance by screening the
ambient fields and it is important to have a velocity field that  
displays large fluctuations at low frequencies. Then, slightly above  
the dynamo onset, on-off intermittency
could be possibly observed.

\subsection{Sodium experiments with weakly constrained flows}

Early experiments have been performed by Lehnert (1957). A swirling flow
has been generated by rotating a disk in a cylindrical vessel containing
$58\, {\rm  l}$ of liquid sodium. Although a self-generated dynamo regime has not
been reached, induction measurements have been performed with an applied
axial field. The generation of a toroidal ($\omega$-effect) and poloidal
field components have been observed.
Another early sodium experiment has been motivated by planetary dynamos for
which tidal forces have been considered as a possible source of power  
driving the
dynamo (Malkus  1968). Precession of a rotating cylinder  filled with  
liquid sodium
has been studied by Gans (1970). The rotation rate was increased
up to $3600\, {\rm  rpm}$ for a precession rate of $50\, {\rm rpm}$. An external
magnetic field amplification has been reported but no self-generation.

Flows in  spherical geometry are currently  investigated in two  
dynamo experiments in the USA. The Wisconsin experiment (Forest {\it et  
al.} 2002, Normberg {\it  et al.} 2006, Spence {\it et al.} 2006) is  
operational: the spherical shell is motionless and the fluid is  
driven by two propellors. This set-up is to be linked with numerical  
work by Dudley and James (1989). No dynamo action has been observed  
so far in the Wisconsin flow. Experiments in a similar geometry but  
at a smaller scale have been performed in Maryland (Peffley {\it et al.}  
2000).

Other sodium experiments, motivated by the context of the geodynamo,  
involve system-wide rotation.
There is a widespread belief that global rotation may assist  
generation by lowering the critical 
Reynolds number.
A spherical Couette flow has been studied in Grenoble. No self-generation has been observed but the effect of the Lorentz force due  
to an external magnetic field has been studied (Nataf {\it et al.} 2006).
A similar instrument is being build in Maryland. A sphere of 3 meter  
in diameter, i.e., much larger than the currently running  
experiments, is being build with another concentric spherical shell  
inside. Both spheres can be put into rotation to investigate the  
effect of global rotation.

\section{The VKS experiment}

\subsection{von K\'arm\'an swirling flows}

The von K\'arm\'an (VK) class of flows consists of flows in which the  
entrainment is performed by coaxial discs (Zandbergen and Dijkstra  
1987). The fluid is enclosed in a cylindrical shell. When the discs  
are operated in counter-rotation, these flows display various qualities  
of interest for a potential dynamo: a strong differential rotation  
and some helicity which are key ingredients for a closed loop  
induction by $\omega$ and $\alpha$ effects. The flow lacks any planar  
symmetry which is also necessary for the $\alpha$ effect. As shown by  
measurements of pressure fluctuations, large vorticity concentrations  
are produced (Fauve {\it et al.} 1993, Abry {\it et al.} 1994) which may also  
act in favour of the amplification of the magnetic field if the  
classical analogy between vorticity and magnetic field production is  
to be believed. The choice of VK flows was thus  
motivated by the hope that the above features will make possible the  
generation a magnetic field by a strongly turbulent flow.

The induction properties of these flows have been performed at  
moderate values of $R_m$ up to 7 in Lyon to evidence the various  
induction steps that may lead to a dynamo (Odier {\it et al.} 1998, 2000,  
Volk {\it et al.} 2006a). With in mind the fact that, in the Riga and  
Karlsruhe dynamos, the observations fitted very well the prediction  
from kinematic dynamo computations, an optimisation procedure of the  
VK flow geometry has been implemented in CEA-Saclay in order to lower the  
threshold of the dynamo instability and to investigate a potential  
influence of the high turbulence level on the bifurcation (Mari\'e {\it et  
al.} 2003, Ravelet {\it et al.} 2005). This procedure underlined in  
particular the importance of the electromagnetic boundary conditions,  
showing that electrically conducting propellors or possibly fluid  
motion behind them can increase $R_{mc}$ by a factor 2 (Stefani {\it et  
al.} 2006).

\subsection{The VKS2 experiment}

\begin{figure}[!htb]
\begin{center}
\includegraphics[width=.6\textwidth]{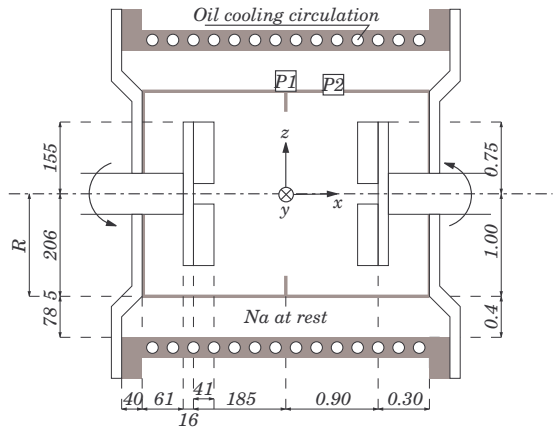}
\caption{Sketch of the VKS experimental set-up. The inner and outer  
cylinders are made of copper (in grey) and the other part in  
stainless steel. The dimensions are given in millimeter on the left  
half of the drawing and normalised by  R on the right half. The 3D  
Hall probe is located either at point P1 in the mid-plane or P2. In  
both cases, the probe is nearly flush with the inner shell. 
(Figure from Monchaux {\it et al.} 2007).}
\label{vks2}
\end{center}
\end{figure}

\begin{figure}[!htb]
\begin{center}
\includegraphics[width=.7\textwidth]{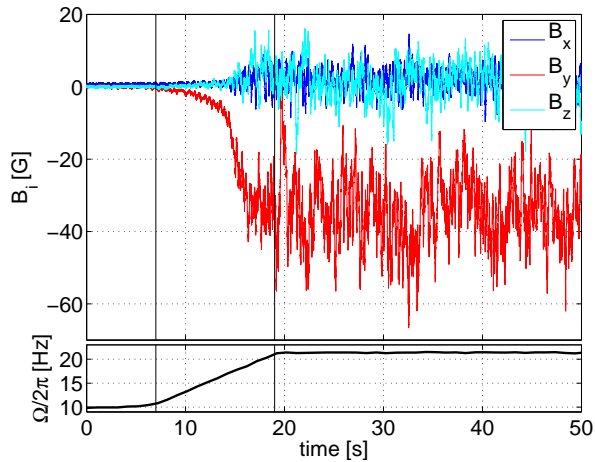}
\caption{Time recording at position $P1$ (fig.~\ref{vks2}) of the  
components of the magnetic field when the rotation frequency $\Omega / 
2\pi$ is increased as displayed by the ramp below ($R_m$ increases  
from $19$ to $40$). (Figure from Monchaux {\it et al.} 2007).}
\label{vkssignal}
\end{center}
\end{figure}

The VKS acronym stands for ``von K\'arm\'an Sodium". The VKS2  
experiment is an evolution of a first design, VKS1 (Bourgoin {\it et al.}  
2002, P\'etr\'elis {\it et al.} 2003) which did not show any dynamo  
action. The changes compared to the VKS1 design are the  
implementation of a cooling system and changes in boundary conditions  
motivated in part by the optimisation procedure discussed above. A  
sketch of the set-up is displayed in figure~\ref{vks2}. The VK flow  
is generated in the inner cylinder of radius $206\,$ {\rm mm} and length $524\,$ {\rm mm}  by two counter rotating discs of radius $154\,$ {\rm mm} and $371\,$ {\rm mm} apart. The 
additions to this base flow are a domain of sodium at rest  
surrounding the VK flow, an annulus in the mid-plane and pure iron  
discs. The sodium at rest has been shown to lower the threshold in  
the kinematic simulations using the mean flow. The annulus has been  
observed to stabilise to some extent the shear layer generated by the  
counter rotation of the discs. These changes in geometry were not  
sufficient to develop the dynamo instability and the last change was  
to use pure iron discs in order to modify the magnetic boundary  
conditions. In addition to the boundary conditions it also de-couples  
to some extent the domain in between the discs from the two domains  
behind the discs, whose flows may not be favourable at least in the  
kinematic simulations. This last configuration enabled the  
observation of a dynamo field as shown in figure~\ref{vkssignal}. As  
the rotation rate of the discs is increased from 10 to 22 {\rm Hz}, one  
observes at the P1 location (see fig. \ref{vks2}) the growth of a  
magnetic field: the azimuthal component acquires a nonzero average value of  
order 40 Gauss with relatively strong fluctuations. The two other  
components display small average values but fluctuate with  
$rms$ values of order 5 gauss. Even though the fluctuation level is  
much higher than in the Karlsruhe or Riga experiment, we call this  
dynamo stationary in the sense that it is not displaying any kind of  
time-periodicity or reversal.

The magnetic Reynolds number in the VKS2 experiment is defined as  
$R_m=K\mu_0\sigma R^2\Omega$ where $R$ is the radius of the cylinder  
and $\Omega$ the rotation rate of the discs. $K=0.6$ is a numerical  
coefficient relating $\Omega R$ to the maximum velocity in the flow.  
With this definition the critical magnetic Reynolds number is close  
to 31 as can be seen from figure~\ref{figdyntot}: the circles   
correspond to the Lundquist number for VKS2 multiplied by a factor $25$.

\subsection{Magnetic energy density of fluid dynamos}

\begin{figure}[!htb]
\begin{center}
\includegraphics[width=.75\textwidth]{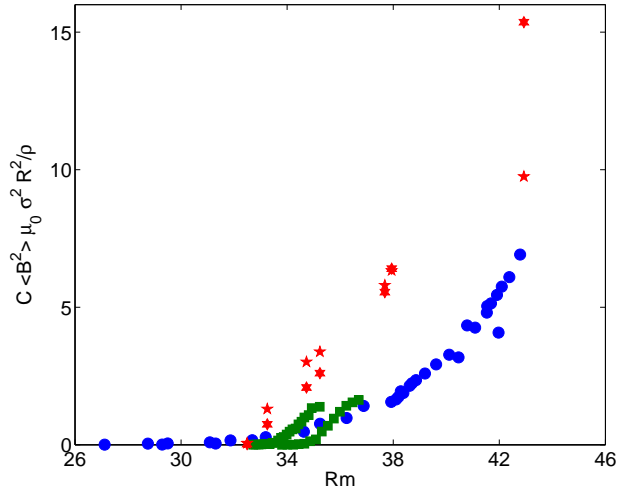}
\caption{Lundquist number as a function of $R_m$: Karlsruhe experiment 
for two ratios of the flow rates ($\blacksquare$), Riga experiment ($ 
\star$) and VKS experiment for various working rotation rates and  
temperatures ($\bullet$). $C$ is a scaling factor used to display all  
curves on a single graph. $C$ is one except for the VKS experiment  
for which $C=25$. We do not expect to need this factor if  
measurements are performed in the bulk of the flow instead at the  
periphery as in Monchaux {\it et al.} 2007.}
\label{figdyntot}
\end{center}
\end{figure}

The dimensionless magnetic energy of the three working dynamos is plotted in figure~\ref{figdyntot} as a function of the magnetic Reynolds number. 
The striking feature is that all  
three dynamos bifurcate at critical $R_m$ in between 30 and 35. Even  
though this may be pure coincidence due to a choice of definitions  
for $R_m$, it nevertheless means that critical values are really of  
the same order of magnitude.  

The saturation values of the three  
dynamos look very similar but again care has  
to be taken here as the Lundquist number is computed by using only a  
few localised measurements and not by integrating over the whole  
volume. This may cause the magnetic energy to be relatively badly  
estimated depending on the measurement point, even if one expects the  
scaling properties to be conserved. For instance, the ad-hoc factor $C=25$ used for the VKS dynamo can be justified because measurements are performed at the boundary of the flow and thus underestimate the magnetic field intensity. 

In any  case the laminar scaling would predict magnetic energy $10^{5}$ too  
small such that for all three dynamos, the  
observations are compatible with the turbulent scaling of the  
saturation of the magnetic energy.

For $R_m>R_{mc}$, the increase of the VKS2 Lundquist number seems of higher order than linear whereas the two other dynamos show an increase compatible with a linear trend. Whether this behaviour is reminiscent  
of some anomalous scaling is a difficult question to answer. The fact  
that the flow displays a much higher turbulence level than the two  
other dynamos may be a justification for anomalous scaling (see  
discussion below) but some care has to be taken in the interpretation  
of the data. In particular, the presence of the iron discs renders  
the bifurcation imperfect because of the remanence of the  
magnetisation in the discs (see discussion below). The apparent nonlinear scaling may be only due to the rather small range of available  
values for $R_m-R_{mc}$ which may not allow to see the actual  
scaling. Also because of the imperfection, the apparent $R_{mc}$ may  
appear lower than the actual one.

\section{Further comments and questions about the VKS experiment}

\subsection{The effect of iron disks}

The addition of the iron discs enabled  the growth of the dynamo field. 
Their effect if first to modify the magnetic boundary  
conditions. For example, when the iron is still in  
the linear regime with permeability $\mu_0 \mu_r$ (i.e., the magnetisation is less that its saturated value), the ferromagnetic metal acts as a shield that prevents the  
field lines to go across the discs. Field lines are refracted when they  penetrate the disc: the normal component of the magnetic field is continuous at the surface but the tangential one is increased by a factor $\mu_r$ in the discs.  
This may de-couple the main flow region in between the discs from the  
motion of the sodium behind the discs which has been shown by  
kinematic simulations based on the time averaged flow to increase the  
threshold of the kinematic dynamo.

The magnetisation vector is also likely to be parallel to the plane  
of the discs in the same way as an elongated rod is more easily  
magnetised along its axis. One possibility that agrees with the axial  
symmetry of the experiment is that the field lines of magnetisation  
would be loops centred on the cylinder axis. This would act in favour  
of magnetic field lines with the shape of loops parallel to the discs  
at least in their vicinity. Such loops with the same orientation as  
the magnetisation would be stabilised by the presence of the  
magnetised disc. Next to the disc, the flow is strongly outwards  
because of the  centrifugal pumping. A material loop of small radius  
close to the disc would be stretched to a larger radius (see the  
sketch in figure~\ref{figvksalphaomega2}). Thus a magnetic field loop  
of the same shape would be amplified by the stretching (Fauve and P\'etr\'elis 2003). The vicinity  of the discs acts as an amplifier for the toroidal component  of the magnetic field.

\begin{figure}[!htb]
\begin{center}
\includegraphics[width=.8\textwidth]{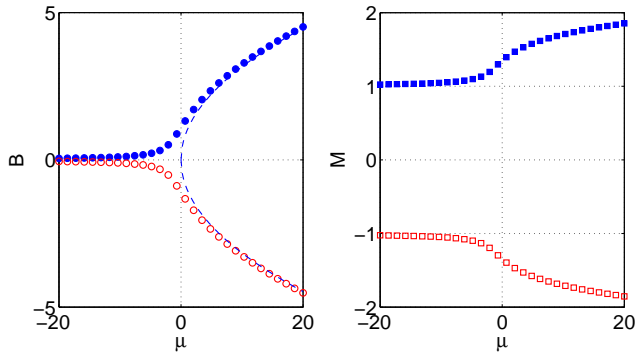}
\caption{Stable solutions of the phenomenological model (\ref{mod1}),  
(\ref{mod2}). On the left is the solution for $B$ and on the right  
the one for $M$. Open and full symbols correspond to the same pair of  
solution $(B,M)$. The dashed line correspond to the classical  
bifurcation without coupling $B^2=\mu$.}
\label{mod}
\end{center}
\end{figure}

As stated before, one consequence of  
the magnetisation of the discs is that the bifurcation is imperfect.  
The imperfection is of a different nature than the one in  
the Karlsruhe experiment. In the latter case, the imperfection was  
due to the Earth's magnetic field and it was possible to explore the  
second branch of solutions by imposing an additional external field  
as shown in figure~\ref{karlsruhebif}. Here the situation is  
different: the magnetisation adds an additional contribution to  
the magnetic field but also the latter can modify the  
magnetisation. As iron has a low coercive force, the magnetisation  
can be flipped by a relatively small magnetic field. Let us  
illustrate this by a simple phenomenological model involving two  
coupled amplitude equations:
\begin{eqnarray}
\partial_t B &=& \mu B -B^3 + M\,, \label{mod1}\\
\partial_t M&=&M-M^3+B\,. \label{mod2}
\end{eqnarray}
The $B$ variable is the analogue of the amplitude of the magnetic field and displays an instability for positive values of $\mu$ without the coupling to the second variable. $M$ is analogous to the magnetisation. The solution $M=0$ is unstable even for $B=0$ as the thermodynamically stable state is  the saturated one and iron has a very narrow hysteretic cycle and low coercive force. Both variables are positively coupled as the presence  
of $B$ induces some change of $M$ due to the susceptibility and  
conversely any nonzero $M$ causes a contribution to $B$. The nonzero  stable solutions are displayed in figure~\ref{mod} (for $\mu>1$ other  solutions exist but these are not relevant for the discussion here).  
One can see the imperfect bifurcation for $B$ but also that both  
the  positive and negative solutions display the same imperfection due to the coupling with the $M$ variable. Indeed $M$ and $B$ always have the same sign because the low coercive force of iron causes  
the flip of the magnetisation as soon as the magnetic field changes  its
sign. This very simple model illustrates the fact that contrary to  
the Karlsruhe dynamo, both branches of the imperfect bifurcation  
display the same feature. Applying an external field may force the  
dynamo to choose one specific branch but the shape would be the same in any case.

Note that the $B=0$, $M=0$ state is always linearly unstable. A more realistic description should incorporate space in order to describe magnetic domains and the metastability of the $M=0$ state. However, the aim of this model is just to illustrate that the sign of the imperfection depends on the one of the bifurcating magnetic field and thus is not prescribed by an external broken symmetry. 

\subsection{A possible dynamo mechanism for the VK flow}

Several predictions have been already presented for the VKS dynamo. They are all based on the mean flow alone;  the neutral mode has been found dipolar with its axis in the mid-plane between the propellors, i.e. an equatorial dipole (Mari\'e { \it et al.} 2003, Bourgoin {\it et al.} 2004,  Ravelet {\it et al.} 2005, Stefani {\it et al.} 2006). 

The geometry of the experimentally observed magnetic field  
differs from these predictions. We note that the mean flow being axisymmetric, it cannot generate an axisymmetric magnetic field according to Cowling's theorem. This constraint does not exist for the mean magnetic field generated by the full flow. We give below another possible mechanism that takes into account the helical structure of  
the radially expelled fluid within two neighbour blades which is averaged out when the mean flow is computed. 

\begin{figure}[!htb]
\begin{center}
\includegraphics[width=.40\textwidth]{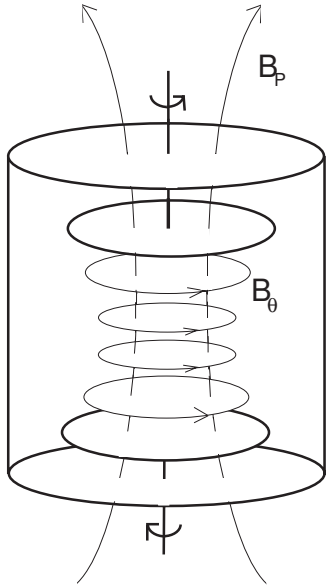}
\caption{A possible $\alpha-\omega$ mechanism for the VKS experiment  
using iron discs. Field loops are amplified by the outwards flow  
close to the discs. The fluid ejected by the discs is also helical  
due to the differential rotation and can convert the toroidal  
component $B_{\theta}$ into a poloidal component $B_P$ 
by $\alpha$ effect. The  
differential rotation converts also very efficiently the poloidal  
component into toroidal by $\omega$ effect. The relative signs of 
$B_{\theta}$ and $B_P$ depend on the sign of the helicity.}
\label{figvksalphaomega2}
\end{center}
\end{figure}

The VK flows are very efficient at converting the poloidal components  
into toroidal one by the $\omega$ induction process: axial field lines are twisted by the strong differential rotation and this  
induces a toroidal component (Odier {\it et al.} 1998). To get a loop back  
reaction from toroidal to poloidal, one can invoke an $\alpha$ effect  
localised close to  the discs: the fluid ejected by the discs is  
strongly helical because of the shear induced by the differential  
rotation. This is the ingredient required for the $\alpha$ effect to  
occur. In this way, an axial field can be induced in the cylinder as  
drawn schematically in figure~\ref{figvksalphaomega2}. 

If one measures the magnetic field on the cylinder, the poloidal  
component is most likely weak as the currents that generate it  
are axisymmetric currents localised in the flow.  On the contrary, the toroidal component  should be stronger than the poloidal one and more likely to be even  stronger when the measurement point is closer to the discs as  observed in position P2. The observations are compatible with this scenario but a deeper investigation of the dynamo field should be performed to validate it.

In any case, an axisymmetric mean magnetic field cannot be generated by the VKS mean flow because of Cowling's theorem. Thus, according to the definition given in section 4, the  
VKS experiment with counter-rotating disks generates a turbulent dynamo.

\section{Open problems and other possible dynamo experiments}

\subsection{Anomalous scaling laws for the magnetic energy density?}

As said in paragraph $3.2$, experimental dynamos operate with liquid  
metals for which $P_m < 10^{-5}$ and neglecting this dimensionless  
number in the expression of the energy density leads to the large $Re$ scaling 
(\ref{largerescaling}). This  
seems reasonable because the magnetic field is dissipated below the  
Ohmic scale $L/R_m^{3/4}$ which is much larger than the Kolmogorov  
scale $L/Re^{3/4}$  at small $P_m$. It is thus unlikely that the  
magnetic energy density depends on kinematic viscosity in this limit.  
However, one should keep in mind that there exist situations where a  
dimensionless parameter cannot be neglected even when it becomes much  
smaller (or larger) than the others. In these infrequent situations,  
the problem is said to be self-similar of the second kind if the  
dependence on this parameter is a power law (Barenblatt  1996). This  
assumption would give an additional dependence $P_m^{\alpha}$ for the  
magnetic energy.

We now want to discuss another ``nonclassical" effect of the same  
class that may result from the presence of strong turbulent  
fluctuations of the velocity field. The magnetic field being forced  
by many different scales, one may expect a situation similar to  
critical phenomena for equilibrium phase transitions where large  
scale thermal fluctuations must be taken into account close to  the  
critical point. This often leads to an ``anomalous" behaviour of the  
order parameter in the vicinity of the transition, where it follows a  
power law with a critical exponent that differs from the mean field  
prediction. Although thermal fluctuations may in principle also  
affect the amplitude of neutral modes in the vicinity of instability  
thresholds, it has been shown that this anomalous behaviour, if it exists, would be limited to a very small range and thus not  
detectable for most hydrodynamic instabilities (Hohenberg and Swift 1992). Various hydrodynamical instabilities have been studied using liquid crystals (Rehberg{\it  et al.} 1991) or in the vicinity of the liquid-vapour critical point (Fauve {\it et al.} 1992, Oh and Ahlers 2003) in order to try to enhance the effect of thermal fluctuations.
Although some effects have been observed, to the best of our knowledge, no anomalous dependence  
of the amplitude of unstable modes above threshold has been reported. Only mean field exponents, i.e., rational power laws given by simple  
symmetry arguments have been measured so far. For a supercritical  
instability, this assumption led to (\ref{turbulentscaling}), i.e., $f 
(R_m,\,P_m)\propto \sqrt{(R_m-R_{mc})}$, thus a mean field exponent  
$1/2$.

As said above, in phase transitions departure from mean-field theory  
occurs when the thermal fluctuations cannot be neglected. This is  
expressed by the Ginzburg criterion that compares the amplitude of the order parameter predicted by mean field theory to the effect of  
thermal fluctuations (Ginzburg 1960). Using a similar criterion, but  
taking into account the kinetic energy $E_F$ of turbulent velocity  
fluctuations instead of $kT$,  would give an interval range above the  
dynamo threshold where an anomalous scaling can be expected. It may be wrong to compare macroscopic turbulent fluctuations to temperature and estimating the number of relevant modes is difficult. However, if we assume that the mean field undergoes a pitchfork bifurcation  and that the dependence of its amplitude in space involves a Laplacian to leading order,  dimensional arguments lead to 
\begin{equation}
\frac{R_{m} - R_{mc}}{R_{mc}} < \left[\frac{\mu_0 E_F}{B_0^2 \xi_0^3} 
\right]^2,
\end{equation}
where $B_0$ is the pre-factor of the magnetic field above threshold in  
mean field theory and $\xi_0$ the one of its correlation length. Inserting the large $Re$ scaling (\ref {turbulentscaling}) for $B_0$, and taking into account that  
fluctuations are of the same order of magnitude as the mean flow, we  
obtain
\begin{equation}
\frac{R_{m} - R_{mc}}{R_{mc}} <  \left[\frac{L}{ \xi_0}\right] ^6,
\end{equation}
that can be of order one even if $\xi_0 \simeq L$.
We thus expect that a critical region in $R_m$ with an anomalous  
scaling of the mean amplitude of the magnetic field can be easily  
observed. This can be understood since the energy of the fluctuations  
is significant compared to the one needed to reverse the magnetic  
field slightly above the dynamo threshold. 

The above estimate of the  
range in $R_{m} - R_{mc}$ where an anomalous behaviour can be  
expected  depends of course on the nature of the bifurcation. For  
instance, if an equatorial mean field were generated, we expect the  
effect of fluctuations to be stronger than for an axisymmetric mean  
field because the orientation of the dipole can easily rotate under  
the influence of fluctuations. 
This is reminiscent of the sensitivity of critical behaviour on the broken symmetry of the ordered phase: fluctuations have a deeper effect when a continuous symmetry is broken. For instance the ordered phase is destroyed by the thermal fluctuations for a two dimensional XY spin model because the ordered phase breaks the axisymmetry of the problem. On the contrary, for a two dimensional  Ising model, the ordered phase is axisymmetric and is not destroyed by the fluctuations (Goldenfeld 1992).

We can argue that velocity fluctuations act in a multiplicative way  
whereas thermal fluctuations in phase transitions are additive. We  
note that in the presence of Earth's magnetic field, velocity  
fluctuations also involve additive forcing. In addition, even in the  
absence of any ambient field, the small scale fluctuations of the magnetic field are forced by an additive term resulting from the interaction between the small scale fluctuations of the velocity and the mean magnetic field generated by the dynamo.  Differences from the mean field behaviour are expected as soon as the   fluctuations of the magnetic field are of the same order as its mean  
value. Together with the hypothesis of self similarity of the second  
kind, the magnetic energy would then behave as
\begin{equation}
\langle\frac{B^2}{\mu_0}\rangle=\rho V^2 P_m^{\alpha} (R_m-R_{mc})^ 
{\beta}\,.
\label{eqenergy}
\end{equation}
with $\beta \neq 1/2$. The results of the VKS experiment cannot presently be used to test
this type of behaviour because of the imperfection due to the iron discs close to threshold.
Thus, it is necessary to increase its maximum magnetic Reynolds number or to find
a more efficient flow in order to obtain a turbulent dynamo without using iron discs.

\subsection{An optimised $\alpha$-$\omega$ dynamo?}

\begin{figure}[htb]
\begin{center}
\includegraphics[width=.40\textwidth]{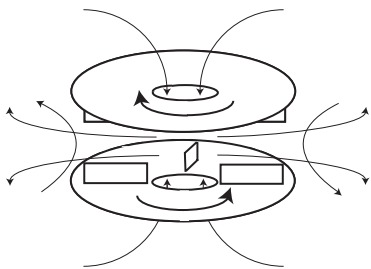}
\caption{An optimised $\alpha$-$\omega$ dynamo experiment.}
\label{figalphaomega}
\end{center}
\end{figure}

If one is to believe the induction scenario proposed above, then it  
may be possible to improve further the efficiency of the effects  
possibly involved in the VKS dynamo. We propose the set-up described  
in figure~\ref{figalphaomega}: a VK flow is driven by two coaxial  
discs with a narrow gap. A very strong shear is thus created  
between these two impellers. The discs are fitted with blades  
resulting in a strong radial outward flow that expels the fluid. The  
discs centres are hollowed: the fluid expelled radially loops back  
into the gap by these openings. The radial velocity  and  the  
azimuthal shear result in a strongly helical flow. Therefore the flow  
displays two ingredients described above: a strong helicity and a  
strong differential rotation that drive respectively an $\alpha$  
effect and an $\omega$ effect. With this modified set-up, one avoids  
the central region of the VKS experiment where the flow is radially  
inbound which tends to dampen the toroidal component of the flow. The  
expected unstable mode is qualitatively similar to the one described  
above for VKS, roughly axisymmetric and made of two parts: a poloidal  
field with same axis as the discs and an azimuthal field. The  
converging region is farther from the amplification region and is  
partially shielded if iron discs are used so that one expects this flow to  
be more efficient and to possibly lead to a lower critical magnetic  
Reynolds number. If the gain of efficiency due to the absence of radially
inward flow in the mid-plane is large enough, one may also observe 
self-generation without using iron discs.

\subsection{Non axisymmetric dynamos}

All currently working dynamos are based on axisymmetric flows. The
time-averaged velocity field of the Riga and VKS dynamos are 
invariant by rotation around the axis of their propellors. For the large
scale magnetic field generated in the Karlsruhe experiment, the
$\alpha$-effect created by the cellular flow is invariant by rotation  
around the axial direction of the flow.

Symmetries usually do not favour dynamo action. There is no $\alpha$-effect 
if the flow displays any planar symmetry. Anti-dynamo theorems are related
to the symmetries of the flow or of the magnetic field.

A non-axisymmetric flow can be generated by driving a fluid with a set of propellors located at positions not aligned along their axis of  
rotation.
A gallium-experiment based on such flow is currently studied at ENS-Paris (Berhanu, Mordant and Fauve 2007). It is driven by two pairs of propellors.  
Depending on their pitch and on the sign of their angular velocity, different flows that  
result from interacting vortices generated by the propellors are observed.  Such  
flows relieve the constraint of axisymmetry and may be efficient for induction
or dynamo action.

In addition, the large scale vortices interact, thus leading to fluctuations of the
flow at low frequencies. This is probably important in order to generate  
dynamos that are strongly affected by the flow fluctuations. 
Low frequency  fluctuations are indeed required in order to have on-off intermittency 
in low dimensional dynamical systems (Auma\^ \i tre {\it et al.}  2005). Similarly,
these fluctuations might favour the occurrence of anomalous exponents for
the magnetic energy of turbulent dynamos.

Finally we point out that a set of vortices generally leads to chaotic
Lagrangian trajectories. This is an important ingredient for fast
dynamo action (Childress and Gilbert 1995).  Although there is no  
hope to reach very large values of $Rm$ in an experimental dynamo,  
studying the growth rate of the magnetic field at moderate $Rm$  
in a flow displaying Lagrangian chaos at large scale can provide 
useful information.

\section{Concluding remarks}

The dynamo threshold and the geometry of the generated magnetic field in the Karlsruhe and Riga experiments were correctly predicted  by taking into account the mean flow alone. 
This observation is supported by an argument showing that due to the low level of fluctuations, the shift in threshold has to be of order two at least in the amplitude of turbulent fluctuations. On the contrary, the geometry of the magnetic field generated in the VKS experiment cannot be generated by the mean flow alone. Large scale turbulent fluctuations are of the same order of magnitude as the mean flow because the flow is much less constrained. Although this needs to be confirmed with a detailed experimental study, we suggest a dynamo mechanism based on a $\alpha-\omega$ process that implies a non stationary part of the VK flow, suggesting that the observed dynamo is turbulent (in the meaning used in this article). 

The three fluid dynamo experiments show that the magnetic energy density in the saturated regime above threshold is such that the Lundquist number is of the order of the distance to criticality. This results from the large value of the kinetic Reynolds number. 

The Karlsruhe and Riga experiments as well as the VKS experiment with propellors in exact counter-rotation, have not displayed secondary bifurcations generating large scale dynamics of the magnetic field.
However, it should be noted that for propellors driven at different rotation frequencies, a great variety of dynamical regimes (stationary, oscillatory, intermittent) as well as reversals of the magnetic field have been observed in the VKS experiment (Berhanu {\it et al.} 2007). It is likely that this results from the presence of competing instability modes that can be also observed in less turbulent dynamos by tuning two different control parameters.  The effect of the fully turbulent character of the VKS2 experiment on these dynamical regimes deserves new experiments.

\bigskip

\noindent
We acknowledge Dr. F. Stefani and Dr. Robert Stieglitz for providing data and figures related to the Riga and Karlsruhe experiments (figures 1-7, 10). We thank all the participants of the VKS collaboration with whom experiments related to figures 8-10 have been performed. 

\vfill
\eject


\section{References}

\smallskip
\noindent
Abry, P.,  Fauve, S., Flandrin, P. and Laroche, C.,  Analysis of  
pressure fluctuations in swirling turbulent flows, {\it J. Physique II}, 1994, {\bf 4},  
725-733.

\smallskip
\noindent
Auma\^ \i tre, S.,  P\'etr\'elis, F. and Mallick, K., Low-- 
frequency noise controls on--off intermittency of bifurcating  
systems, {\it Phys. Rev. Lett.} 2005 {\bf 95},  064101.

\smallskip
\noindent
Auma\^ \i tre, S., Mallick, K., Petrelis, F., Effects of the low  
frequencies of Noise on On-off intermittency, {\it Journal of Statistical  
Physics}, 2006, {\bf 123}, 909-927.

\smallskip
\noindent
Barenblatt, G. I., {\it Scaling, Self-similarity, and Intermediate  
Asymptotics}, 1996 (Cambridge University Press).

\smallskip
\noindent
Bassom and Gilbert, Nonlinear equilibration of a dynamo in a smooth helical flow, {\it J. Fluid Mech.}, 1997, {\bf 343}, 375-406.

\smallskip
\noindent
Berhanu, M., Monchaux, R., Fauve, S., Mordant, N., P\'etr\'elis, F., Chiffaudel, A., Daviaud, F., Dubrulle, B., Mari\'e, L., Ravelet, F., Bourgoin, M., Odier, Ph., Pinton, J.-F., Volk, R., Magnetic field reversals in an experimental turbulent dynamo, {\it Europhys. Lett.}, 2007, {\bf 77}, 59001.	

\smallskip
\noindent
Berhanu, M., Mordant, N. and Fauve, S., Ecoulement turbulent dans  
un cylindre: haut nombre de Reynolds et fluctuations \`a basse fr \'equence, 
(Rencontres du non lin\'eaire, 2007).

\smallskip
\noindent
Berthet, R., Petrossian, A., Residori,  S., Roman B. and Fauve, S.
Effect of multiplicative noise on parametric instabilities,   
{\it Physica  D} 2003 {\bf 174}, 84-99.

\smallskip
\noindent
Bourgoin, M. , Mari\'e, L., P\'etr\'elis, F., Gasquet, C., Guigon,  
A., Luciani, J. B ., Moulin, M., Namer, F., Burgete, J., Chiffaudel,  
A., Daviaud, F.,  Fauve, S.,  Odier, P. and Pinton, J. F.,   
Magnetohydrodynamics measurements in the von K{\'a}rm{\'a}n sodium  
experiment,  {\it Phys. Fluids}, 2002, {\bf 14}, 3046-3058.

\smallskip
\noindent
Bourgoin, M. , Odier, P.  Pinton, J. F. and Ricard, Y., An  
iterative study of time independent induction effects in  
magnetohydrodynamics, {\it Phys. Fluids}, 2004, {\bf 16}, 2529-2547.

\smallskip
\noindent
Busse, F. H., M\"uller, U., Stieglitz, R. and Tilgner, A.,
A two-scale homogeneous dynamo, and extended analytical model and  
an experimental demonstration under development,  
{\it Magnetohydrodynamics}, 1996, {\bf 32}, 235-248.

\smallskip
\noindent
Busse, F. H., A model of the geodynamo, {\it Geophys. J. R. Astr.  
Soc.}, 1975,  {\bf 42}, 437-459.

\smallskip
\noindent
Cattaneo, F. and Hughes, D. W., Dynamo action in a rotating  
convective layer, {\it J. Fluid Mech.}, 2006, {\bf 553}, 401-418.

\smallskip
\noindent
Childress, S. and Gilbert, A. D.  {\it Stretch, Twist, Fold: the fast  
dynamo},  1995 (Berlin, Heidelberg: Springer Verlag).

\smallskip
\noindent
Childress, S. and Soward, A. M., Convection driven hydromagnetic  
dynamo, {\it Phys. Rev. Lett.}, 1972,
{\bf 29}, 837-839.

\smallskip
\noindent
Dudley, M. L. and James, R. W., Time-dependent kinematic dynamos  
with stationary flows , {\it Proc. R. Soc. London A}, 1989, {\bf 425}, 407- 429.

\smallskip
\noindent
Fauve, S., Kumar, K., Laroche, C., Beysens, D. and Garrabos, Y.,
Parametric instability of a liquid-vapor interface close to the critical point, {\it Phys. Rev. Lett.}, 1992,  {\bf 68}, 3160.
 
\smallskip
\noindent
Fauve, S., Laroche, C. and Castaing, B., Pressure fluctuations in  
swirling turbulent flows, {\it J. Physique II}, 1993, {\bf 3}, 271-278.

\smallskip
\noindent
Fauve, S. and Lathrop, D. P.,  Laboratory experiments on liquid  
metal dynamos and liquid metal MHD turbulence. In {\it Fluid Dynamics and  
Dynamos in Astrophysics and Geophysics}, A. Soward et al. (Eds), pp.  
393-425, 2005, (CRC Press).

\smallskip
\noindent
Fauve, S. and P\'etr\'elis, F.,  The dynamo effect, in  
{\it Peyresq Lectures on Nonlinear Phenomena}, Vol. II,  J-A  
Sepulchre (Ed), pp. 1-64 2003 (Singapore: World Scientific).

\smallskip
\noindent
Fauve, S. and P\'etr\'elis, F., Scaling laws of turbulent dynamos,
{\it C. R. Physique}, 2007, {\bf 8}, 87-92.

\smallskip
\noindent
Forest, C. B., Bayliss, R. A., Kendrick, R. D., Nornberg, M. D.,  
O'Connell, R. and Spence, E. J.,  Hydrodynamic and numerical  
modeling of a spherical homogeneous dynamo experiment,  
{\it Magnetohydrodynamics}, 2002, {\bf 38}, 107-120.

\smallskip
\noindent
Frick, P., Denisov, S., Khripchenko, S, Noskov, V., Sokoloff, D. and  
Stepanov, R., A nonstationary dynamo experiment in a braked torus. In {\it Dynamo and Dynamics, a Mathematical Challenge}, P.  
Chossat et al. (Eds), pp. 1-8,  2001 (Dordrecht, Kluwer Academic Publishers).

\smallskip
\noindent
Gailitis, A., Lielausis, O., Platacis, E.,  Dement'ev, S.,
Cifersons, A., Gerbeth, G., Gundrum, T., Stefani, F., Christen M. and  
Will, G., Magnetic field saturation in the Riga dynamo experiment, 
{\it Phys. Rev. Lett.}, 2001, {\bf 86}, 3024-3027.

\smallskip
\noindent
Gailitis, A., Lielausis, O., Platacis, E., A., Gerbeth, G., Stefani,  
F.,  Laboratory experiments on hydromagnetic dynamos, {\it Rev. Mod.  
Phys.}, 2002, {\bf 74}, 973990.

\smallskip
\noindent
Gailitis, A., Lielausis, O., Platacis, E., Gerbeth, G., Stefani, F., 
The Riga dynamo experiment,  {\it Surveys in Geophysics}, 2003, {\bf 24}, 247-267.

\smallskip
\noindent
Galloway, D. J. and Proctor, M. R. E.,  Numerical calculations of  
fast dynamos in smooth velocity fields with realistic diffusion,  
{\it Nature (London)}, 1992,  {\bf 356}, 691-693.

\smallskip
\noindent
Gans, R. F., On hydromagnetic precession in a cylinder, {\it J. Fluid
Mech.}, 1970,  {\bf 45}, 111-130.

\smallskip
\noindent
Gilbert, A. D. and Sulem, P. L., On inverse cascades in alpha  
effect dynamos, {\it GAFD}, 1990,  {\bf 51}, 243-261.

\smallskip
\noindent
Ginzburg, V. L., Some remarks on phase transitions of the second kind and the  
microscopic theory of ferroelectric materials, {\it Soviet Phys. Solid  
Sate}, 1960,  {\bf 2}, 1824-1834.

\smallskip
\noindent
Goldenfeld, N., {\it Lectures on Phase Transitions and the Renormalization Group},  1985 
(Addison-Wesley).

\smallskip
\noindent
Hohenberg, P. C.  and Swift, J. B., Effects of additive noise at the onset of Rayleigh-B\'enard convection, {\it Phys. Rev. A}, 1992, {\bf 46}, 4773-4785.

\smallskip
\noindent
Hoyng, P., Helicity fluctuations in mean field theory: an explanation for the variability of the solar cycle?, {\it Astron. Astrophys.}, 1993, {\bf 272}, 321-339.

\smallskip
\noindent
Hughes, D. W.  and Cattaneo, F., The $\alpha$-effect in rotating  
convection: size matters, submitted to {\it J. Fluid Mech.}, 2007.

\smallskip
\noindent
Iroshnikov, P. S., Turbulence of a conducting fluid in a strong  
magnetic field, {\it Soviet Astron.}, 1963, {\bf 7}, 566-571.

\smallskip
\noindent
John, T.,  Stannarius, R.  and Behn U., On-Off Intermittency in  
stochastically driven electrohydrodynamic convection in nematics,  
{\it Phys. Rev. Lett.}, 1999, {\bf 83}, 749-752.

\smallskip
\noindent
Kabashima, S., Kogure, S., Kawakubo, T. and Okada T.,
Oscillatory-to-nonoscillatory transition due to external noise in a  
parametric oscillator, {\it J. Appl. Phys.}, 1979,  {\bf 50}, 6296-6302.

\smallskip
\noindent
Knobloch, E., private communication (2007).

\smallskip
\noindent
Kraichnan, R. H.,
Inertial-range spectrum of hydromagnetic turbulence, {\it Phys. Fluids}  
{\bf 8}, 1965, 1385-1387.

\smallskip
\noindent
Krause, F. and R\"adler, K.-H., {\it Mean Field
Magnetohydrodynamics and Dynamo Theory}, 1980
(New-York: Pergamon Press).

\smallskip
\noindent
Larmor, J., How could a rotating body such as the
sun become a magnet?, Rep. $87^{\rm th}$ Meeting Brit. Assoc. Adv.
Sci., Bornemouth, Sept. 9-13, 1919, pp. 159-160, 1919 (London: John Murray).

\smallskip
\noindent
Laval J. P., Blaineau, P., Leprovost, N., Dubrulle, B. and Daviaud,  
F., Influence of turbulence on the dynamo threshold, {\it Phys. Rev.  
Lett.}, 2006, {\bf 96}, 204503.

\smallskip
\noindent
Lehnert, B., An experiment on axisymmetric flow of liquid sodium in
a magnetic field, {\it Arkiv f\"or Fysik}, 1957,  {\bf 13}, 109-116.

\smallskip
\noindent
L\'eorat, J., Lallemand, P., Guermond, J. L. and Plunian, F., Dynamo  
action, between numerical experiments and liquid sodium devices. In  
{\it Dynamo and Dynamics, a Mathematical Challenge}, P. Chossat et  
al. (Eds), pp. 25-33,  2001 (Dordrecht, Kluwer Academic Publishers).

\smallskip
\noindent
L\"ucke, M.  and Schank, F.,  Response to parametric modulation  
near an instability, {\it Phys. Rev. Lett.}, 1985,  {\bf 54},
1465-1468.

\smallskip
\noindent
Malkus, W. V. R,  Precession of the Earth as the cause of  
geomagnetism, {\it Science}, 1968,  {\bf 160}, 259-264.

\smallskip
\noindent
Mari\'e, L.,  Burguete, J.,  Daviaud, F.  and L\'eorat, J.,  
Numerical study of homogeneous dynamo based on experimental von K\'arm\'an type flows,
{\it Eur. Phys. J. B}, 2003, {\bf 33}, 469-485.

\smallskip
\noindent
Moffatt, H. K., The amplification of a weak applied magnetic field  
by turbulence in fluids of moderate conductivity, {\it J. Fluid Mech.}, 1961, 
{\bf 11}, 625-635.

\smallskip
\noindent
Moffatt, H. K., {\it Magnetic Field Generation in Electrically  
Conducting Fluids}, 1978 (Cambridge University Press).

\smallskip
\noindent
Monchaux, R., Berhanu, M.,  Bourgoin, M.,  Moulin, M.,  Odier, Ph.,  
Pinton, J.-F., Volk, R., Fauve, S., Mordant, N., P\'etr\'elis, F.,   
Chiffaudel, A., Daviaud, F., Dubrulle, B., Gasquet, C. and Mari\'e,  
L., Generation of magnetic field by dynamo action in a turbulent  
flow of liquid sodium, {\it Phys. Rev. Lett.}, 2007, {\bf 98}, 044502.

\smallskip
\noindent
M\"uller U., Stieglitz R. and Busse F. H., On the sensitivity of  
dynamo action to the system's magnetic diffusivity,
{\it Phys. Fluids}, 2004,  {\bf 16}, L87-90.

\smallskip
\noindent
Nataf, H. C., Alboussi\`ere, T., Brito, D., Cardin, P., Gagni\`ere,  
Jault, D., Masson, J.-P. and Schmitt, D., Experimental study of  
super-rotation in a magnetostrophic spherical Couette flow, {\it GAFD},
2006,  {\bf 100}, 281- 298.

\smallskip
\noindent
Nornberg M. D., Spence, E. J., Kendrick, R. D., Jacobson, C. M.  and  
Forest, C. B., Intermittent magnetic field excitation by a  
turbulent flow of liquid sodium,  {\it Phys. Rev. Lett.}, 2006, {\bf 97}, 044503.

\smallskip
\noindent
Nu\~nez, A., P\'etr\'elis, F.  and Fauve, S., Saturation of a
Ponomarenko type fluid dynamo. In {\it Dynamo and Dynamics, a  
Mathematical Challenge}, P. Chossat et al. (Eds), pp. 67-74,  2001 (Dordrecht, Kluwer
Academic Publishers).

\smallskip
\noindent
Odier, P., Pinton, J. F. and Fauve, S.,  Advection of a magnetic  
field by a turbulent swirling flow, {\it Phys. Rev. E}, 1998, {\bf 58},   
7397-7401.

\smallskip
\noindent
Odier, P., Pinton J.-F. and S. Fauve,   Magnetic induction by  
coherent vortex motion, {\it Eur. Phys. J. B}, 2000, {\bf 16}, 373-378.

\smallskip
\noindent
Oh, J. and Ahlers, G., Thermal noise effect on the transition to Rayleigh-B\'enard convection,  2003, {\bf 91}, 094501.

\smallskip
\noindent
Parker, E. N., The occasional reversal of the geomagnetic field, {\it Astrophys. J.}, 1969, {\bf 158}, 815-827.
\smallskip
\noindent
Peffley, N. L. , Cawthorne, A. B.  and  Lathrop, D. P., Toward a  
self-generating magnetic dynamo: The role of turbulence, {\it Phys. Rev.  
E}, 2000, {\bf 61}, 5287-5294.

\smallskip
\noindent
P\'etr\'elis, F., {\it PhD Thesis},  2002 (Universit\'e Paris 6).

\smallskip
\noindent
P\'etr\'elis, F. and Auma\^{\i}tre,  S., Intermittency at the edge  
of a stochastically inhibited
pattern--forming instability, {\it Eur. Phys. J. B}, 2003, {\bf 34}, 281-284.

\smallskip
\noindent
P\'etr\'elis, F.,   Auma\^ \i tre, S.  and Fauve, S.,  Effects of phase  
noise on the Faraday instability, {\it Phys. Rev. Lett.}, 2005, {\bf 94}, 070603.

\smallskip
\noindent
P\'etr\'elis,  F.  and Auma\^ \i tre ,  S., Modification of  
instability processes by multiplicative noises, {\it Euro. Phys. J. B.}, 2006, {\bf 51}, 357-362.

\smallskip
\noindent
P\'etr\'elis, F. and Fauve, S.,  Saturation of the magnetic field  
above the dynamo threshold, {\it Eur. Phys. J. B} , 2001, {\bf 22}, 273-276.

\smallskip
\noindent
P\'etr\'elis, F. and Fauve, S., Inhibition of the dynamo effect by phase  
fluctuations, {\it Europhys. Lett.}, 2006,  {\bf 76}, 602-608.

\smallskip
\noindent
P\'etr\'elis, F., Bourgoin,  M. Mari\'e, L., Burgete, J., Chiffaudel,  
A., Daviaud, F., Fauve, S., Odier, P. and Pinton, J. F.,  Nonlinear  
magnetic  induction by helical motion in  a liquid sodium turbulent  
flow, {\it Phys. Rev. Lett.}, 2003, {\bf 90}, 174501.

\smallskip
\noindent
Platt, N., Hammel, S. and Heagy, J., Effect of additive noise on on- 
off intermittency, {\it Phys. Rev. Lett.}, 1994,  {\bf 72}, 3498-3501.

\smallskip
\noindent
Platt, N., Spiegel, E.A.  and Tresser, C.,  On-off intermittency: A  
mechanism for bursting, {\it Phys. Rev.
Lett.}, 1993,  {\bf 70}, 279-282, and references therein.

\smallskip
\noindent
Ponomarenko, Yu. B., Theory of the hydromagnetic generator, {\it J.  
Appl. Mech. Tech.
Phys.}, 1973, {\bf 14}, 775-778.

\smallskip
\noindent
Ponty Y. , Mininni, P. D. , Montgomery, D. C., Pinton, J.-F.,  
Politano, H. and Pouquet, A., Numerical Study of Dynamo Action at  
Low Magnetic Prandtl Numbers,  {\it Phys. Rev. Lett.}, 2005, {\bf 94}, 164502; 
Ponty, Y., Mininni, P., Pinton, J. F., Politano, H. and  
Pouquet, A., Dynamo action at low magnetic Prandtl numbers: mean  
flow vs. fully turbulent motion (arXivphysics/0601105), to be  
published in {\it New J. Phys.}, 2007.

\smallskip
\noindent
R\"adler, K.-H., Apstein, E., Rheinhardt, M. and Sch\"uler, M.,
The Karlsruhe dynamo experiment. A mean field approach,
{\it Studia Geophys. Geod.}, 1998, {\bf 42}, 224-231.

\smallskip
\noindent
Ravelet F., Chiffaudel, A., Daviaud, F. and L\'eorat, J., Toward an  
experimental von K\'arm\'an dynamo: Numerical studies for an optimized  
design, {\it Phys. Fluids}, 2005,  {\bf 17}, 117104.

\smallskip
\noindent
Rehberg, I., Rasenat, S., de la Torre Ju\'arez, M., Sch\"opf, W., H\"orner, F., Ahlers, G. and Brand, H. R., Thermally induced hydrodynamic fluctuations below the onset of electroconvection, {\it Phys. Rev. Lett.}, 1991, {\bf 67}, 596-599.

\smallskip
\noindent
Reighard, A. B. and Brown, M. R., Turbulent conductivity  
measurements in a spherical liquid sodium flow, {\it Phys. Rev. Lett.},
2001, {\bf 86}, 2794-2797.

\smallskip
\noindent
Residori, S., Berthet, R., Roman, B. and Fauve, S., Noise induced  
bistability of parametric surface waves, {\it Phys. Rev. Lett.}, 2001, {\bf 88},  
024502.

\smallskip
\noindent
Roberts, G. O.,  Dynamo action of fluid motions with two- 
dimensional periodicity, {\it Phil. Trans. Roy. Soc. London A}, 1972,
{\bf 271}, 411-454.

\smallskip
\noindent
Roberts, P. H.,  Future of geodynamo theory,  {\it GAFD}, 1988,  {\bf 44}, 3-31, 
and references therein.

\smallskip
\noindent
Roberts, P. H., Fundamentals of dynamo theory. In {\it Lectures on Solar and Planetary Dynamos }, pp. 1-57, M. R. E. Proctor and A. D. Gilbert (Eds),  1994 (Cambridge University Press).

\smallskip
\noindent
Ruzmaikin, A. A. and Shukurov A. M.,
Spectrum of the galactic magnetic fields, {\it Astrophysics and Space  
Science}, 1982,  {\bf 82}, 397-407.

\smallskip
\noindent
Soward, A. M., A convection-driven dynamo: I. The weak field case, {\it Phil. Trans. R. Soc. Lond.},
1974,  {\bf A 275}, 611-646.

\smallskip
\noindent
Spence, E. J.  Nornberg M. D., Jacobson, C. M.,  Kendrick, R. D.,    
and Forest, C. B.,  Observation of a turbulence-induced large scale magnetic field, 
{\it Phys. Rev. Lett.}, 2006, {\bf 96}, 055002.

\smallskip
\noindent
Stefani, F, Xu, M., Gerbeth, G., Ravelet, F., Chiffaudel, A.,  
Daviaud, F. and L\'eorat, J., Ambivalent effects of added layers on  
steady kinematic dynamos in cylindrical geometry: application to the  
VKS experiment, {\it Eur. J. Mech. B}, 2006, {\bf 25} 894.

\smallskip
\noindent
Stieglitz, R. and M\"uller,  U., Experimental demonstration of a  
homogeneous two-scale dynamo, {\it Phys. Fluids}, 2001,
 {\bf 13}, 561-564.

\smallskip
\noindent
Stieglitz, R. and M\"uller, U, Experimental demonstration of a
homogeneous two-scale dynamo,
{\it Magnetohydrodynamics}, 2002, {\bf 38}, 27-34.

\smallskip
\noindent
Stratonovich, R.L., {\it Topics in the Theory of Random Noise},  1963
( London: Gordon and Breach).

\smallskip
\noindent
Sweet, D., Ott, E.,  Finn, J., Antonsen, T. M.  and Lathrop, D.,  
Blowout bifurcations and the onset of magnetic activity in  
turbulent dynamos, {\it Phys. Rev. E}, 2001,  {\bf 63},  066211,1-4.

\smallskip
\noindent
Tilgner, A. and Busse, F. H., Saturation mechanism of a
model of the Karlsruhe dynamo. In {\it Dynamo and Dynamics, a  
Mathematical Challenge}, P. Chossat et al. (Eds), pp. 109-116,   2001 (Dordrecht, Kluwer  
Academic Publishers).

\smallskip
\noindent
Verma, M. K., Statistical theory of magnetohydrodynamic turbulence:  
recent results,
{\it Phys. reports}, 2004, {\bf 401}, 229-380.

\smallskip
\noindent
Volk R., Odier P. and Pinton J.-F., Fluctuation of magnetic  
induction in von K\'arm\'an swirling flows, {\it Phys. Fluids}, 2006, {\bf 18},  
085105.

\smallskip
\noindent
Volk, R., Ravelet,  F., Monchaud, R., Berhanu,  M., Chiffaudel, A.,  
Daviaud, F., Odier, P., Pinton, J.-F., Fauve, S., Mordant, N. and  
P\'etr\'elis, F., Transport of magnetic field by a turbulent flow of  
liquid sodium, {\it Phys. Rev. Lett.}, 2006, {\bf 97}, 074501.

\smallskip
\noindent
Zandbergen, P. J. and Dijkstra, D., von K\'arm\'an swirling flows,  
{\it Annu. Rev. Fluid Mech.} {\bf 19}, 1987, 465-491.

\smallskip
\noindent
Zeldovich, Ya. B., Ruzmaikin, A. A.  and Sokoloff, D. D.,
{\it Magnetic Fields in Astrophysics},  1983 (New
York: Gordon and Breach).

\end{document}